\documentclass[10pt,twocolumn,prd,a4paper,preprint,nofootinbib,superscriptaddress]{revtex4}

\usepackage{graphicx}
\def\lb{\linebreak[4]}

\usepackage{wrapfig}

\usepackage{latexsym,epsfig}
\usepackage{amsmath,amssymb,bm}
\usepackage{longtable}

\newcommand{\be}{\begin{equation}}
\newcommand{\ee}{\end{equation}}
\newcommand{\bes}{\begin{subequations}}
\newcommand{\ees}{\end{subequations}}
\newcommand{\bea}{\begin{eqnarray}}
\newcommand{\eea}{\end{eqnarray}}
\newcommand{\bear}{\begin{equation}\begin{array}}
\newcommand{\eear}[1]{\end{array}\label{#1}\end{equation}}
\def\ba{$$\begin{array}}
\def\ea{\end{array}$$}
\def\bra{$\begin{array}}
 \def\era{\end{array}$}

\newcommand{\fr}[2]{\dfrac{{ #1}}{{ #2}}}

\def\vep{{\varepsilon}}
\newcommand{\epe}{\mbox{$e^+e^-\,$}}
\newcommand{\ggam}{\mbox{$\gamma\gamma\,$}}

\newcommand{\ggww}{\mbox{$\gamma\gamma\to W^+W^-\,$}}

\newsavebox{\fmbox}

{\end{list}}
\newcounter{enumct}

\newcommand{\bu}{$\bullet$\ }

\begin{document}

\title{Charge asymmetries in \bm{$\gamma\gamma \to \ell^+\ell^-+
\nu$}{\em's} ($\ell=\mu,\,e$) with polarized photons in the Standard
Model.}

\author{D.~A.~Anipko}
\affiliation{Sobolev Institute of Mathematics and Novosibirsk State University, Novosibirsk, 630090, Russia}

\author{M.~Cannoni}
\affiliation{Universit\`a di Perugia, Dipartimento di Fisica,
Via A.~Pascoli, I-06123, Perugia, Italy}
\affiliation{Istituto Nazionale di Fisica Nucleare, Sezione di Perugia, Via A.~Pascoli, I-06123, Perugia, Italy}

\author{I.~F.~Ginzburg}
\affiliation{Sobolev Institute of Mathematics and Novosibirsk State University, Novosibirsk, 630090, Russia}

\author{K.~A.~Kanishev}
\affiliation{Sobolev Institute of Mathematics and Novosibirsk State University, Novosibirsk, 630090, Russia}
\affiliation{University of Warsaw, 00-681 Warsaw, Poland}

\author{A.~V.~Pak}
\affiliation{Sobolev Institute of Mathematics and Novosibirsk State University, Novosibirsk, 630090, Russia}
\affiliation{Department of Physics, University of Alberta, Edmonton, AB T6G 2G7,  Canada}

\author{O.~Panella}
\affiliation{Istituto Nazionale di Fisica Nucleare, Sezione di Perugia, Via A.~Pascoli, I-06123, Perugia, Italy}

\begin{abstract}

It is shown that in reaction $\gamma\gamma\to\ell^+\ell^-+\nu's$ at
$\sqrt{s}>200$ GeV with polarized photons, large and well observable
differences arise in the distribution of positive and negative
charged leptons ($\ell=\mu^\pm,\,e^\pm$), (charge asymmetry). The
modification due to the contribution of the cascade processes with
intermediate $\tau$-lepton in $\gamma \gamma \to W^\pm\ell^\mp +
\nu's$ reaction is taken into account. This charge asymmetry is
potentially sensitive to effects of physics beyond the standard
model at the anticipated luminosity of  the Photon Collider mode of
the future international linear collider.

\end{abstract}

\date{10/6/2008}
\maketitle

\section{Introduction}
\label{secintr}

The Photon Collider (PC) option of the planned International Linear
Collider (ILC)
(see e.g.~\cite{Tesla},~\cite{Ginzburg1}) will offer a specific window for
the study of new effects in both Standard Model (SM) and New Physics. In
particular, it is expected that the charge asymmetry of leptons,
produced in the collision of {\it neutral but highly polarized
colliding particles} $\gamma\gamma\to \ell^+\ell^- +neutrals$ (where
$\ell=\mu,\,e$), can be a good tool for the discovery of New Physics
effects. With this aim the study of such asymmetry in SM is a
necessary step for both better knowledge of SM and understanding of
background for New Physics effects.

In this paper we study  the SM process, in which {\it neutrals} are
neutrinos and the main (but not single) mechanism for charged lepton
production is given by $\gamma \gamma \to W^+W^-$ process with
subsequent lepton decay of $W$. The latter process, $\sigma(\gamma
\gamma \to WW) Br (W \to \mu\nu) = 8.8$ pb,  will ensure very high
event rate at the anticipated integrated luminosity of ILC (100
fb$^{-1}$), about $10^6$~events per year. The charge asymmetry here
appears due to transformation of initial photon helicity into
distribution of final leptons via P-violating but CP-preserving
leptonic decay of $W$. In the following we consider the particular
case ${\ell=\mu}$ for definiteness. The considered effects are
identical for electrons and muons. So that, absolutely the same
asymmetry will be observed in $e^+\,e^-$, $e^+\,\mu^-$, $\mu^+\,e^-$
distributions. All these contributions should be added for a
complete analysis. This will enhance the value of the cross section
for $\gamma\gamma\to\mu^+\mu^-+\nu's$ from 1.2 to 4.8 pb.

In the main body of the paper we consider the collision of a photon
with helicity $\lambda_1$ moving in the positive direction of the
$z$ axis with a photon of helicity $\lambda_2$ moving in the
opposite direction. This initial state  is denoted as
$\gamma_{\lambda_1}\gamma_{\lambda_2}$ with $\lambda_i=\pm$ (left or
right circular polarization). For example, the initial state with
$\lambda_1=+1$, $\lambda_2=-1$ is written as $\gamma_+\gamma_-$.
With this choice of the positive direction of the $z$ axis we define
the longitudinal momentum $p_\|\equiv p_z$ and the transverse
momentum $p_\bot\equiv \sqrt{p_x^2+p_y^2}$. For definiteness, we
present most of the results  for monochromatic photon beams at
$\sqrt{s}_{\gamma\gamma} = 500$~GeV ($E_\gamma=250$~GeV). The above
definitions will be slightly  modified when discussing  the  effects
due to the non-monochromaticity of photon beams in the future Photon
Collider. We start our numerical calculations with the CompHEP
package~\cite{CompHEP} and then switch to the  CalcHEP
package~\cite{Pukhov} which allows one to take into account the
circular polarization of the initial photons and choose different
random seed numbers for the Monte Carlo (MC) generator which is
necessary for an estimate of the statistical inaccuracy of future
experiments.

The observable final state with $W+\mu$ or two muons  with missing
transverse momentum carried away by neutrinos can appear either via
processes
 \begin{equation}
\gamma\gamma \to W\mu \nu \qquad
(\gamma\gamma \to \mu^+\mu^- \nu_{\mu}
\bar{\nu}_\mu)\, ,
 \label{mainproc}
 \end{equation}
or via {\it cascade processes} like:
 \be
 \begin{array}{cl}
 \ggam\to W^+\bar{\nu}_\tau&\tau^-\\
 &\downarrow\\
 &\mu^-\bar{\nu}_\mu\nu_\tau\,
 \end{array} =W^+\bar{\nu}_\tau\mu^-\bar{\nu}_\mu\nu_\tau
\ee
\be
 \begin{array}{cll}
\ggam\to&\tau^- \bar{\nu}_\tau\nu_\tau &\tau^+\\
 &\downarrow&\downarrow\\
 &\mu^-\bar{\nu}_\mu\nu_\tau&\mu^+\nu_\mu\bar{\nu}_\tau\,
\end{array}=\mu^-\mu^+\bar{\nu}_\mu\nu_\tau\nu_\mu\bar{\nu}_\tau\
 \bar{\nu}_\tau\nu_\tau
, \label{cascadeproc}
 \ee
in which six or eight particles are present in the final state.

To reduce CPU time in the Monte Carlo event generation, $10^6$
events for each channel, we obtain the essential part of the results
for the $\ggam\to W^\pm\mu^\mp +\nu's$ process, not taking into
account issues related to the reconstruction of the $W$. The
analysis of this process allows us to extract the main features of
the effect of interest, i.e. the difference in the distributions of
$\mu^+$ and $\mu^-$ at fixed photon helicities (\emph{global charge
asymmetry}). We show that the  additional diagrams that contribute
to $\gamma\gamma\to \ell^+\ell^- +neutrals$ give negligible
contribution to the cross section and charge asymmetry.

We start with the description in Section~\ref{secgeneral} of the
general features of the effect, neglecting cascade processes. We classify the diagrams contributing
to $\gamma\gamma \to W\mu \nu$ and $\gamma\gamma \to \mu^+\mu^- \nu_{\mu}
\bar{\nu}_\mu$
according to the different topologies and give
an approximate analytical estimate of their relative
impact on the cross section (Sec.~\ref{secdiagr}).
These estimates allow us to present
qualitative explanation of the appearance of the charge asymmetry (Sec.~\ref{secqual}).

We then introduce suitable variables for the description of the {\it
global asymmetry}, i.e. the difference in distributions of $\mu^+$
and $\mu^-$ in the processes $\gamma\gamma\to W^\pm\mu^\pm+\nu$'s or
$\ggam\to\mu^+\mu^-+\nu$'s (Sec.~\ref{secglob}). In this very
section we describe the cuts applied to the observed particles. In
Sec.~\ref{secstat} we discuss a computational method used to
estimate a lower bound on the statistical uncertainty of future
experiments. This estimate is obtained directly by the repeated
Monte Carlo simulations with an anticipated number of events.

Section~\ref{secasWmu} is devoted to detailed description of the
global charge asymmetry of leptons in the process $\ggam\to W\mu\nu$
in  monochromatic $\ggam$ collisions.

The accurate calculation of cascade processes with six or more
particles in the final state is a computationally challenging task
with available software. Since we use CompHEP/CalcHEP packages which
don't fix the helicity of final states while the  discussed effects
strongly depend on the helicity, the direct use of existing software
for tau decay simulation like TAUOLA~\cite{tauola}  is not possible
here. In Sec.~\ref{seccasccontr} we construct reasonable
approximations in the description of cascade processes
\eqref{cascadeproc}.  The detailed analysis of the modification of
momentum distributions allows to find that the inaccuracy
implemented by the mentioned approximation in the final result is
within the estimated statistical uncertainty of future experiments.

In Sec.~\ref{secresult} we discuss the total observable asymmetries.

High energy photons will be produced at the Photon Collider through
Compton back-scattering of laser photons from high energy electron
or (and) positron beams: the photons will not be monochromatic but
will demonstrate an energy and polarization distribution. The high
energy part of this spectrum will mainly include photons with
definite helicity $\lambda_i$ close to $\pm 1$~\cite{Ginzburg1}. We analyse the influence of initial photon
non-monochromaticity on results in Sec.~\ref{non-mon}.

{\it The correlative asymmetry} in $\mu^+$ and $\mu^-$ momenta in
each event of $\gamma\gamma \to \mu^+\mu^- +\nu's$ is expected to
provide more information in the search for effects of physics beyond
the SM. We discuss it in Sec.~\ref{seccorrel}.

We conclude and summarize the obtained results in Sec.~\ref{secdisc}.

Preliminary (and incomplete) parts of this work were reported earlier~\cite{Anipko}.

\section{General features}\label{secgeneral}

The SM cross section of $\gamma \gamma \to W^+W^-$
at center of mass energy greater than $200$ GeV remains almost constant
at the asymptotic value $\sigma \simeq  8\pi\alpha^2 /{M_W^2}\simeq 80$ pb
and practically independent on
photon polarization~\cite{GKPS}, see the formulas in Subsection~\ref{secqual}.
At $\sqrt{s}>200$~GeV this cross
section is more than ten times larger than the cross section of $W$
production in \epe\, mode.
It will ensure very high event rate at the anticipated  luminosity. The
distributions of $W^+$ and $W^-$ bosons in $\gamma \gamma \to
W^+W^-$ process are identical (charge symmetrical distribution),
their polarizations are determined by the polarization of initial
photons. The distribution of muons in subsequent decay of polarized
$W^\pm$
is asymmetrical due to P non-conservation with CP conservation
in the SM.

\subsection{Diagrams}
\label{secdiagr}

In this section we classify all tree level diagrams describing
the process $\gamma \gamma \to W^\pm\mu^\mp \nu$ and
$\gamma \gamma \to \mu^+\mu^- \nu\bar{\nu}$
in classes according to their topology
(a similar classification was given also in Ref.~\cite{Boos}).
For each topology we
give an analytical estimate of
its asymptotic contribution to the total cross section
at $s\gg M_W^2$, identifying
in each group the $2\to 2$  dominant subprocess and assuming for the SM gauge
couplings $g^2\sim g'^2\sim e^2=4\pi\alpha$.
The numerical Monte Carlo results, supporting these estimates, are presented in the next Sections.

The processes $\gamma \gamma \to W^\pm\mu^\mp \nu$ are described by
seven diagrams, which we divide in three classes, shown in
Fig.~\ref{figdiagwmu}:
\begin{figure}[t!]
\begin{center}
\includegraphics[scale=0.6]{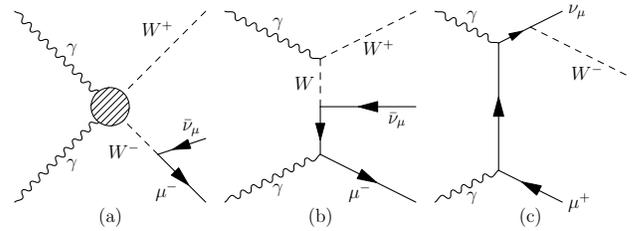}
\caption{Classes of tree level Feynman diagrams contributing
to $\gamma \gamma \to W^+\mu^- \bar\nu$, (a--c). The grey blob in (a)
represents diagrams with $W$ exchange with trilinear $\gamma WW$ coupling
and the diagram with quartic $\gamma\gamma WW$ coupling.}
\label{figdiagwmu}
\end{center}
\end{figure}
\begin{itemize}
\item[(a)]
Three double--resonant diagrams (DRD) of
Fig.~\ref{figdiagwmu}(a) describe $WW$ pair production with
subsequent decay. Their contribution to the total cross section is
{$ \sigma_d\! \sim\! \sigma_{\gamma\gamma\to
WW}Br(W\!\to\!\mu\nu)\! \sim\! (\alpha^2/M_W^2)Br(W\!\to\!\mu\nu)$.}
\item[(b)]
Two single--resonant diagrams (SRDW) of  Fig.~\ref{figdiagwmu}(b)
with $W$ exchange in $t$--channel contribute to the total cross
section $\sigma_{s}\! \sim\! \alpha\sigma_{\gamma\gamma\to WW} \sim
(\alpha^3/M_W^2)$. The relation between this contribution and
DRD contribution is $\sigma_{s}/\sigma_d\sim\alpha/Br(W\to\mu\nu)$.
\item[(c)]
Two single resonant diagrams (SRD$\mu$) with
lepton exchange in $t$--channel (gauge boson bremsstrahlung),
Fig.~\ref{figdiagwmu}(c). The contribution to the total cross
section is $\sigma_{s\mu}\sim \alpha\sigma_{\gamma\gamma\to\mu\mu}
\sim(\alpha^3/s)$, therefore
$\sigma_{s\mu}/\sigma_d \sim [\alpha/Br(W\to\mu\nu)](M_W^2/s).$
\end{itemize}

\begin{figure}[t!]
\begin{center}
\includegraphics[scale=0.6]{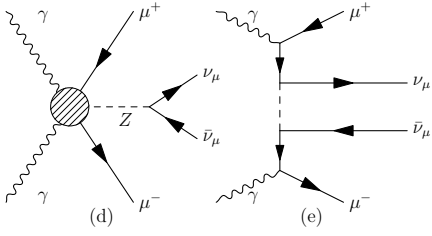}
\caption{
Additional tree level Feynman
diagrams contributing to $\gamma \gamma \to
\mu^+\mu^-\nu\bar\nu$: the grey blob in (d) represents diagrams with $\mu\mu$
fusion to $Z$ and diagrams with $Z$ radiated by an external $\mu$ line.}
\label{figdiagmumu}
\end{center}
\end{figure}

The process $\gamma \gamma \to \mu\mu\nu\bar\nu$ is described
by the diagrams Fig.~\ref{figdiagwmu}
with the addition of lines describing the
$W\to\mu\nu$ decay and permutations of external fermion lines
(with the same estimates as
above) and two additional types of diagrams shown in
Fig.~\ref{figdiagmumu}.

\begin{itemize} 
\item[(d)]
Six diagrams with radiation of $Z$ boson in the process
$\gamma\gamma\to\mu^+\mu^-$, Fig.~\ref{figdiagwmu}(d). The
asymptotic contribution is $\sigma_Z \sim
\alpha\sigma_{\gamma\gamma\to\mu\mu}\alpha Br(Z\to\nu\bar{\nu}) \sim
(\alpha^3/s)Br(Z\to\nu\bar{\nu})$.
\item[(e)]
Two multi-peripheral non-resonant diagrams
Fig.~\ref{figdiagwmu}(e) with $\sigma_n\sim\alpha^4/M_W^2$.
\end{itemize}

\subsection{Qualitative picture}\label{secqual}

The above analysis shows that the bulk of the cross section is given
by the diagrams containing the process \ggww\ with subsequent decay
of $W$ bosons to leptons Fig.~\ref{figdiagwmu}(a), DRD
diagrams.
\begin{widetext}
Denoting by $p_{\bot}$ W's transverse momentum, the
\ggww\ differential cross section can be written as~\cite{GKPS}:
 \bear{c}
d\sigma = d\sigma^{np}+\lambda_1 \lambda_2 d\tau^a\,, \quad
\sigma_W=\fr{8\pi\alpha^2}{M_W^2}\,,\quad x=\fr{4M_W^2}{s}\,,\quad
dt =
-dp_\bot^2\sqrt{\fr{s}{s-4(p_\bot^2+M_W^2)}}\,,  \\[4mm]
d\sigma^{np}=\sigma_W \left[\fr {(16+3x^2)M_W^2}{32(p_{\bot}^2+M_W^2)^2}
-\fr{(3+8x)x}{32(p_{\bot}^2 +M_W^2)}
+\fr{3x^2}{64M_W^2}\right]dt,\\[2mm]
 d\tau^a=\sigma_W \left [-\fr
{xM_W^2}{2(p_{\bot}^2 +M_W^2)^2}+\fr{(3+8x)x}{32(p_{\bot}^2 +M_W^2)}
-\fr{3x^2}{64M_W^2}\right]dt\,,
 \eear{gewnudifptr}
with total cross section ($v=\sqrt{1-x\,}$),
 \be
 \sigma
=\sigma_{W}v \cdot \left \{1+\fr {3x}{16}+\fr {3x^{2}}{16}-
(1-\fr {x}{2}) \fr{3x^2}{16v}\log\fr{1+v}{1-v}
  +\fr{x\lambda_1\lambda_2}{16}
\cdot\left[-19+ \fr{(8-5x)}{v}\log\fr{1+v}{1-v}\right]\right\}\,.
 \ee
\end{widetext}
From the analysis of these equations we can see that at
$\sqrt{s}>200$ GeV the \ggww\ differential cross section practically
does not depend on photon polarizations, as for standard QED
process, total cross section is practically energy independent, at
the plateau value $\sigma_W={8\pi\alpha^2}/{M_W^2}$. Moreover, the
$W$'s are produced mainly in the forward and backward directions,
with average transverse momentum $\sim M_W$ (distribution $\propto
1/(p_\bot^2+M_W^2)^2$).

As shown in Ref.~\cite{BBB}, in this process we have an approximate
helicity conservation. For $p_\bot=0$, the helicity of $W^\pm$
moving in the positive direction of $z$ axis is
$\lambda_{W_1}=\lambda_1$, irrespective of the charge of the $W$;
the same holds for the $W$ moving in the opposite direction:
$\lambda_{W_2}=\lambda_2$. These identities do not hold for $p_\bot
\neq 0$ and become less and less accurate with increasing values of
$p_\bot$. Since in our process the cross section is concentrated at
small values of $p_\bot$, we have an approximate helicity
conservation: $\lambda_{W_1}\approx \lambda_1$, and
$\lambda_{W_2}\approx \lambda_2$,  both for $W^+$ and $W^-$.

Now we can qualitatively understand the origin of charge
asymmetries. Let $z'$--axis be directed along $W$ three--momentum
and $\varepsilon\approx M_W/2$ and $p_{z'}$ be the energy and the
longitudinal momentum of $\mu$ in the $W$ rest frame. It is easy to
calculate that the distribution of muons from the decay of $W$ with
charge $e=\pm 1$ and helicity $\lambda=\pm 1$ in its rest frame is
$\propto(\varepsilon -e\lambda p_{z'})^2$ (the transverse momenta of
muons are distributed roughly isotropically relative to $W$ momentum
within the interval $p_\bot<m_W/2$). In other words, the
distribution of muons from $W^\pm$ decay has a peak along $W$
momentum if the $e \cdot \lambda_W=-1$ and opposite to $W$ momentum
if $e\cdot \lambda_W=+1$. These distributions are boosted  to the
distributions in the $\gamma\gamma$ collision frame. For example,
for a collision of photons in a $\gamma_-\gamma_-$ initial state,
the $\mu^-$ are distributed around the upper value of their
longitudinal momentum (in forward and backward direction), while the
$\mu^+$ are concentrated near the zero value of their longitudinal
momentum. At the same time, this boost makes the distribution in
$p_\bot$ wider in the first case and narrower in the second case.

\subsection{Cuts}\label{seccuts}

It is natural to expect that the relative size of New Physics
effects will be enhanced with the growth of transverse momenta of
observed particles. This is the main reason why we study the
dependence of observed effects on the cut in $p_\bot $. Namely, we
impose cuts on the transverse momenta of observed charged particles,
$p_{\bot \mu}^c$ and on the scattering angle
\bear{c}
p_\bot>p_{\bot \mu}^c\,,\\[1mm]
\theta_0<\theta<\pi-\theta_0\,,\qquad \theta_0=p_{\bot \mu}^c/2E\,.
\eear{eqcuts} The cut \eqref{eqcuts} is applied to each observed
particle and to the total transverse momentum for the sum of momenta
of all observed particles; the cut for escape angle is applied to
all the observed particles. We consider the dependence of all
studied quantities on the $p_{\bot \mu}^c$ up to $p_{\bot
\mu}^c=140$~GeV. This cut also mimics  limitations from the detector
in the future experiment.

These simultaneous cuts allow to eliminate many backgrounds (since
charged particle(s) with missing transverse momentum greater than
$p_{\bot \mu}^c$ should have the escape angle greater than $2p_{\bot
\mu}^c/\sqrt{s}>\theta_0$).  In particular, all pure QED and QCD
processes are eliminated by these cuts, since they cannot provide
large missing transverse momentum.

If it is not otherwise specified, in the following
we set $p_{\bot\mu}^c=10$~GeV, $\theta_0=20$~mrad,
and  monochromatic photon beams at
$\sqrt{s}_{\gamma\gamma} = 500$~GeV ($E_\gamma=250$~GeV).

The effect of cuts with non-monochromatic photons is studied in sect.~\ref{non-mon}.

\subsection{Variables for description of global asymmetry}
\label{secglob}

The global asymmetry variables are described by the difference in
distributions of $\mu^+$ and $\mu^-$ in the processes
$\gamma\gamma\to W^+\mu^-+\nu$'s and $\gamma\gamma\to
W^-\mu^++\nu$'s (or in $\ggam\to\mu^+\mu^-+\nu$'s). For
definiteness, we calculate all quantities only for {\it the case
when negatively charged particle ($W^-$ or $\mu^-$) is in the
forward hemisphere ($ p_\|>0$)}. In the study of the dependence on
$p_{\bot\mu}^c$, we will label all the quantities by the argument
$(p_{\bot\mu}^c)$.

A suitable measure for the {\it longitudinal} ($\Delta_L$) and {\it transverse} ($\Delta_T$) charge asymmetries are
the relative differences  of corresponding momenta distributions for negative and positive muons:
\be
\Delta_L=\fr{ \int p_\|^- d\sigma -\int p_\|^+ d\sigma} { \int
p_\|^- d\sigma +\int p_\|^+ d\sigma} \,,\quad \Delta_T=\fr{ \int
p_\bot^- d\sigma -\int p_\bot^+ d\sigma}{ \int p_\bot^- d\sigma
+\int p_\bot^+ d\sigma} \,.\label{eqPLT}
\ee
It is useful to define also mean values of
longitudinal $p_\|^\mp$ and transverse $p_\bot^\mp$ momenta of
$\mu^-$ or $\mu^+$
 \be
P_{L}^\pm=\fr{\int p_\|^\pm d\sigma}{ E_{\gamma max}\int d\sigma}
\,,\qquad P_{T}^\pm= \fr{\int p_\bot^\pm d\sigma}{ E_{\gamma
max}\int d\sigma}\,.\label{meanp}
 \ee
(These definitions are written in the form which is useful for
non-monochromatic case as well.)

Due to CP symmetry of the SM,
  \begin{eqnarray}
&&  d\sigma_{--}(p_{\mu^+},
p_{\mu^-})=d\sigma_{++}(p_{\mu^-},p_{\mu^+})\cr
&&d\sigma_{+\,-}(p_{\mu^+}, p_{\mu^-})=d\sigma_{-\,+}(p_{\mu^-},
p_{\mu^+})\,.\label{CPsimdsigma}
  \end{eqnarray}
(Here subscripts $+$ and $-$ at cross section label initial photon
helicities). \footnote{One can write the differential distribution in the reaction $\ggam\to\mu^+\mu^-+\nu$'s as
 \be
\fr{d\sigma}{d^3p_{\mu^+}d^3p_{\mu^-}}=
A+B\lambda_1+C\lambda_2+D\lambda_1\lambda_2
 \label{extrarel}
 \ee
That is another form of eq.~\eqref{CPsimdsigma} with
$\int Bd^3p_{\mu^+}d^3p_{\mu^-}=0$, $\int Cd^3p_{\mu^+}d^3p_{\mu^-}=0$. The  weak dependence of the  cross section on the photon polarization  means that $\int Dd^3p_{\mu^+}d^3p_{\mu^-}\ll\int Ad^3p_{\mu^+}d^3p_{\mu^-}$. Our subsequent analysis based on momentum distributions shows that on average $|D|\sim|B|\sim |B|\sim |A|$. In the following paragraphs and sections  we will not make use of the  form in Eq.~\eqref{extrarel}. }.
In particular, the distributions of $\mu^- $ and
$\mu^+$ in the forward hemisphere for $\gamma_+\gamma_-$ collision
reproduce the distributions of $\mu^+$ and $\mu^-$ in the backward
hemisphere. Therefore, in all the cases the asymmetries $\Delta_L$
(determined by Eq.~\eqref{eqPLT}) change signs in each hemisphere
when the helicity changes to opposite. For the $\gamma_-\gamma_+$
collisions these asymmetries in forward and backward hemispheres
have opposite signs. These symmetries break if any CP-violating
interaction is present.

Total cross sections of the processes $\gamma\gamma\to
W^+\mu^-+\nu$'s and $\gamma\gamma\to W^-\mu^++\nu$'s coincide at
each initial photon polarization. However, in accordance with above
discussed qualitative picture, applied cuts reduce these cross
section in different way. So that, it is useful to define relative
value of this difference in dependence on cut variable,
\be
\Delta\sigma(p_{\bot\mu}^c)=\fr{\left(\int d\sigma(W^-\mu^+)- \int
d\sigma(W^+\mu^-)\right)_{(p_\bot>p_{\bot\mu}^c)}} {\left(\int
d\sigma(W^-\mu^+)+ \int
d\sigma(W^+\mu^-)\right)_{(p_\bot>p_{\bot\mu}^c)}}\,,\label{crsecdif}
\ee
and the fraction of the total
cross section left by the cut in $p_{\bot\mu}^c$:
 \be
\sigma^\pm(p_{\bot\mu}^c)= 
{\int
d\sigma(W^\mp\mu^\pm)|_{(p_\bot>p_{\bot\mu}^c)}}.
 \label{relcrsec}
\ee

\subsection{Estimate of statistical uncertainties}
\label{secstat}

MC calculations simulate an experiment and have some statistical
uncertainty $\delta_{MC}$. This uncertainty  value for the integral
characteristics like~\eqref{eqPLT} cannot be predicted simply from
general reasons. To find this uncertainty we repeated our MC
calculation with anticipated $10^6$ number of events five times for
different random number inputs for MC generator. Additionally we
consider as an independent input the set of observations  obtained
by simultaneous change $\lambda_1,\,\lambda_2\to
-\lambda_1,\,-\lambda_2$, $\mu^-\leftrightarrow \mu^+$ (this change
should not change distributions due to CP conservation in SM), in
whole it corresponds ten repetitions of "MC experiment" in a sum.
These sets of data were an input for standard generation of Monte
Carlo inaccuracies $\delta_{MC}$.

Since the adaptive MC is used with CalcHEP, it is natural to expect
that the statistical uncertainty of future real experiment
$\delta_{exp}^{stat}\ge\delta_{MC}$. Therefore, below we  omit
subscript MC, having in mind that our numbers give estimate for
statistical uncertainty from below.

\section{Global asymmetries in the main process
{\boldmath $\ggam\to W^\pm \mu^\mp \nu $}.
Monochromatic case}
\label{secasWmu}
\begin{figure}[b!]
\begin{center}
\includegraphics[scale=0.23]{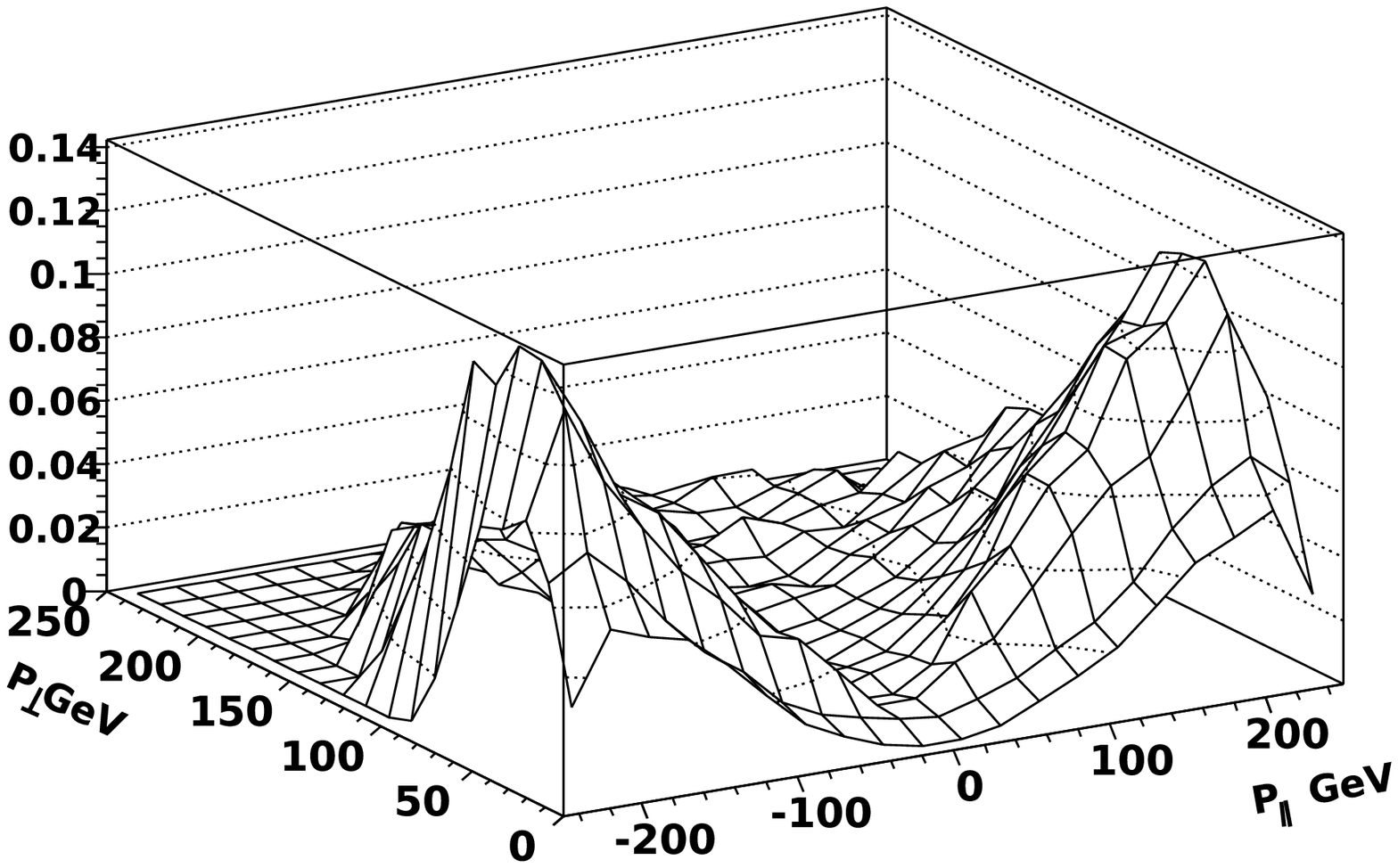}
\includegraphics[scale=0.23]{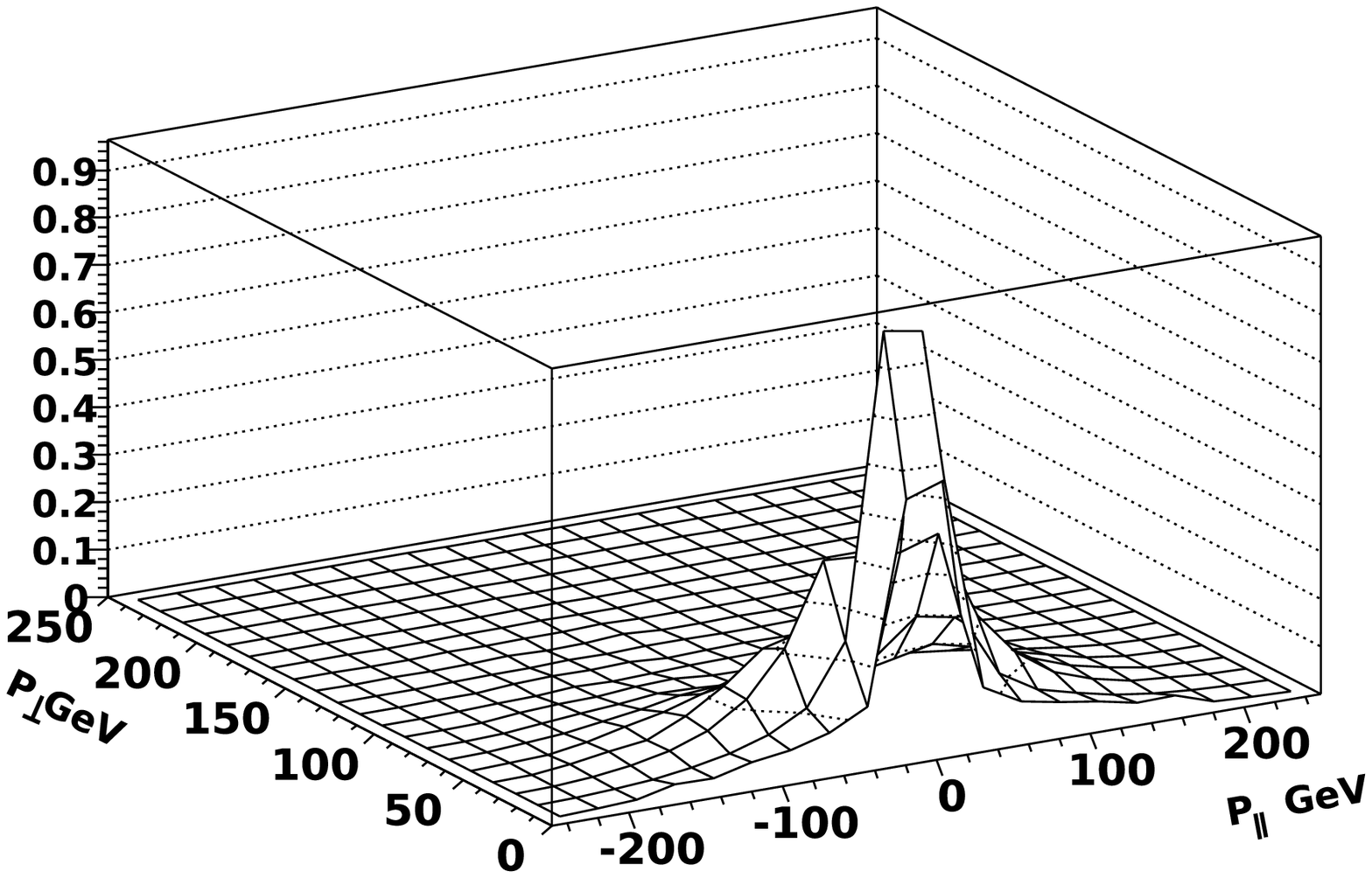}\\
\includegraphics[scale=0.23]{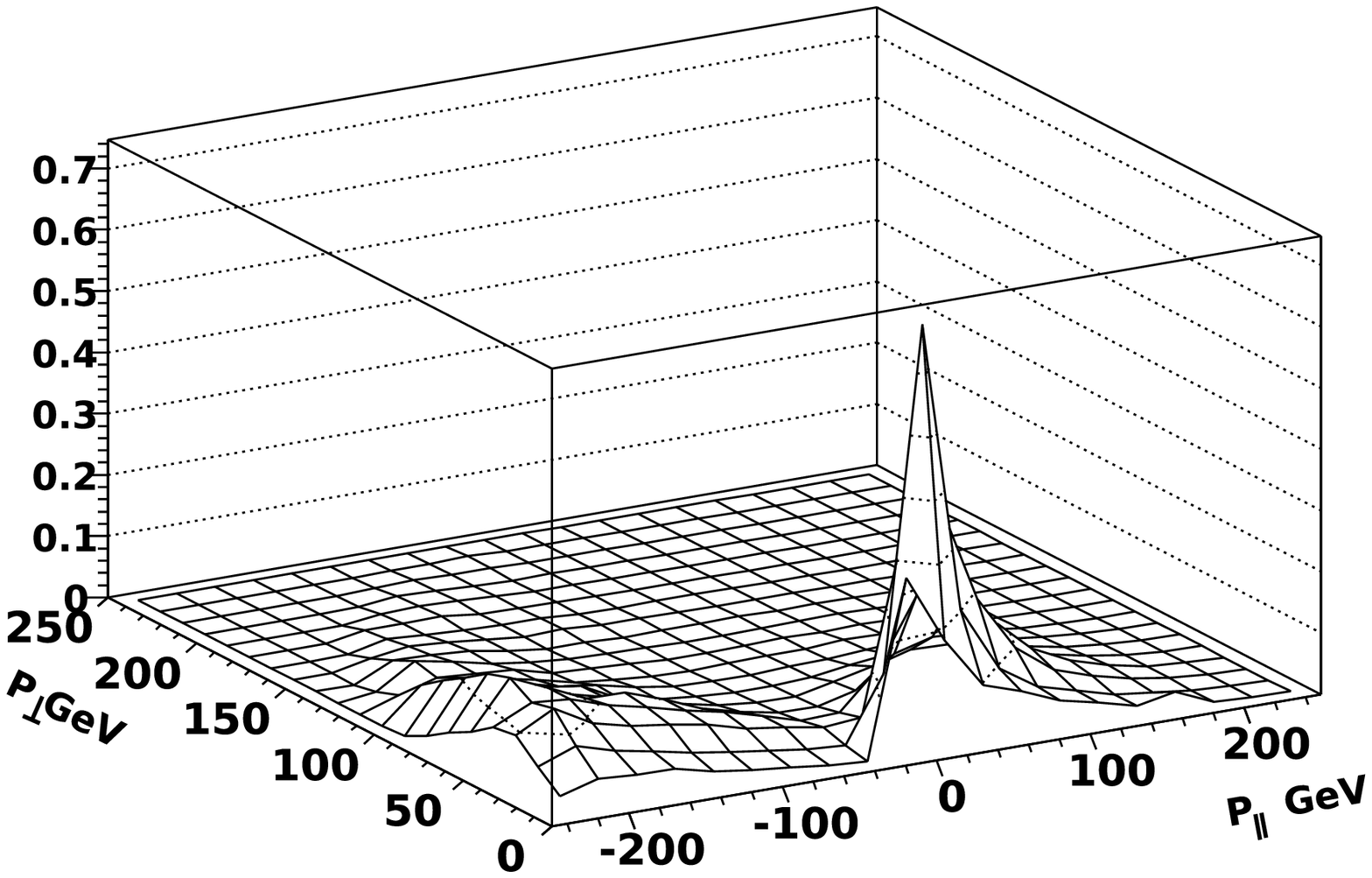}
\includegraphics[scale=0.23]{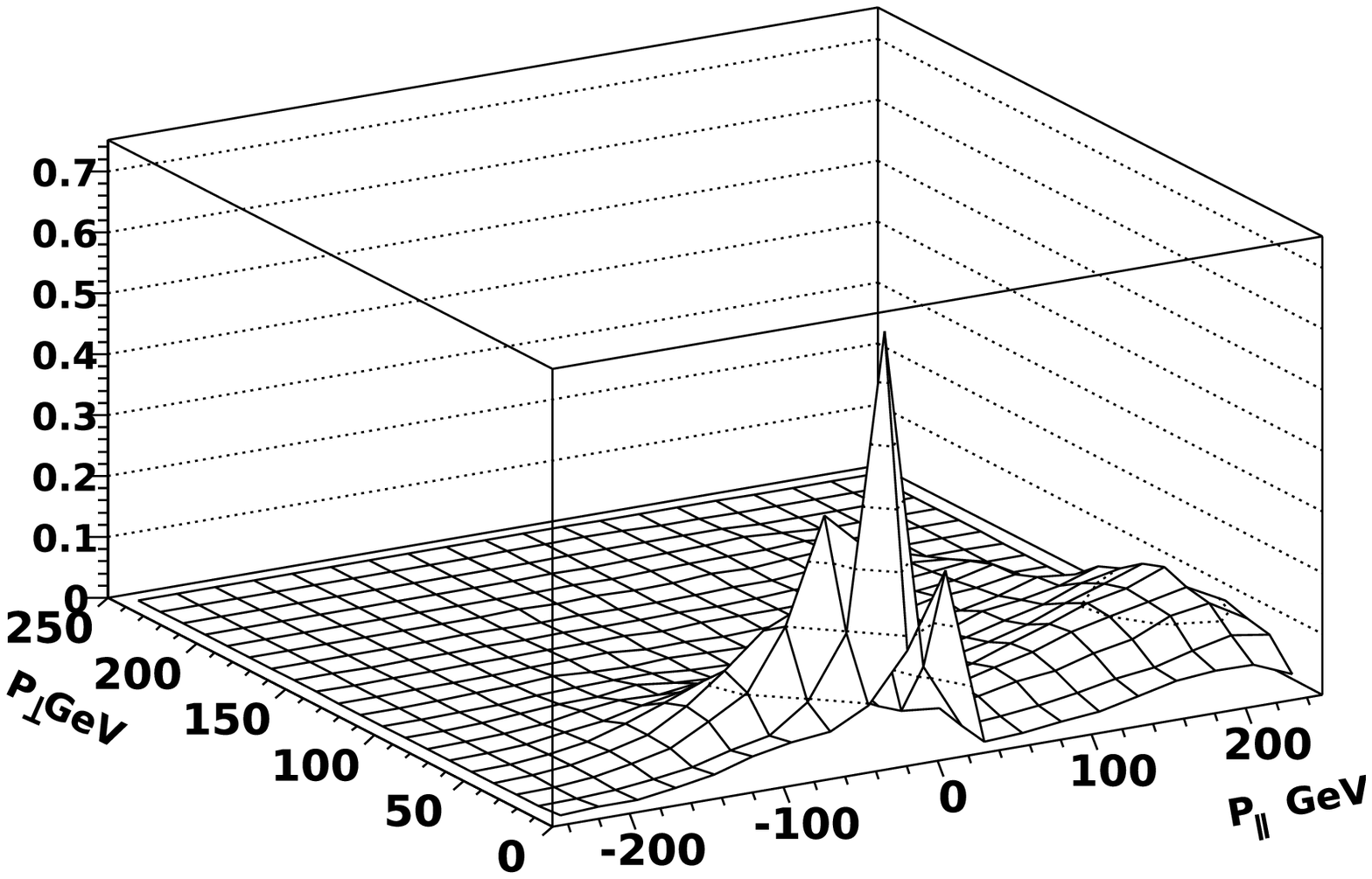}
\caption{Muon distribution in $\gamma_-\gamma_-\to W\mu\nu$
(upper plots) and in $\gamma_+\gamma_-\to W\mu\nu$ (lower
plots), left column $\mu^-$, right column $\mu^+$.}
\label{asymm_no_tau+-}
\end{center}
\end{figure}

\begin{center}
\begin{table*}[t!]
\begin{tabular}{||c|c||c|c|c|c|c|c|c|c|c|c|c|c||}\hline\hline
$p_{\bot\mu}^c$&$\gamma_{\lambda_1}\gamma_{\lambda_2}$ &$P_L^-$ &
$\delta P_L^-$& $P_L^+$ &$\delta P_L^+$& $\Delta_L$& $\delta
\Delta_L$&$P_T^-$ & $\delta P_T^-$& $P_T^+$ &$\delta P_T^+$&
$\Delta_T$& $\delta \Delta_T$\cr \hline
  10& $\,
\gamma_{-}\gamma_{-}\,$& $\, 0.606\,$ &$\, 0.29\% \,$ & $\, 0.201\,
$ & $\,0.55\,$\% &$\, +0.501\, $ & $\, 0.57\% \,$ &$\, 0.333\, $&
$\, 0.61\%\, $ &$\, 0.159\,$ &$\, 0.28\%\,$ &$\, +0.355\, $ &
$\,0.44\%\, $\cr \cline{2-14}
 &$\, \gamma_{+}\gamma_{-}\,$& $\, 0.223\,$ &$\, 0.74\% \,$ & $\, 0.609\,
$ & $\,0.19\,$\% &$\, -0.463\, $ & $\, 0.47\% \,$ &$\, 0.164\, $
&$\, 0.08\%\, $ &$\, 0.262,$ &$\, 0.31\%\,$ &$\, -0.231\, $ &
$\,0.76\%\, $\cr\hline\hline
 40& $\,
\gamma_{-}\gamma_{-}\,$& $\, 0.593\,$ &$\, 0.39\% \,$ & $\, 0.273\,
$ & $\,0.20\,$\% &$\,+0.370\, $ & $\, 0.47\% \,$ &$\, 0.378\, $& $\,
0.64\%\, $ &$\, 0.241\,$ &$\, 0.62\%\,$ &$\, +0.222\, $ &
$\,1.07\%\, $\cr \cline{2-14}
 &$\, \gamma_{+}\gamma_{-}\,$& $\, 0.296\,$ &$\, 0.64\% \,$ & $\, 0.637\,
$ & $\,0.25\,$\% &$\, -0.366\, $ & $\, 0.66\% \,$ &$\, 0.239\, $
&$\, 0.28\%\, $ &$\, 0.319\,$ &$\, 0.25\%\,$ &$\, -0.143\, $ &
$\,0.31\%\, $\cr\hline\hline
 140& $\,
\gamma_{-}\gamma_{-}\,$& $\, 0.402\,$ &$\, 0.68\% \,$ & $\, 0.242\,
$ & $\,0.14\,$\% &$\, +0.249\, $ & $\, 1.45\% \,$ &$\, 0.697\, $&
$\, 0.11\%\, $ &$\, 0.621\,$ &$\, 0.04\%\,$ &$\, +0.057\, $ &
$\,0.95\%\, $\cr \cline{2-14}
 &$\, \gamma_{+}\gamma_{-}\,$& $\, 0.253\,$ &$\, 0.81\% \,$ & $\, 0.489\,
$ & $\,0.27\,$\% &$\, -0.318\, $ & $\, 1.33\% \,$ &$\, 0.672\, $
&$\, 0.12\%\, $ &$\, 0.660\,$ &$\, 0.05\%\,$ &$\, -0.009\, $ & $\,
5.75\%\, $\cr\hline\hline
\end{tabular}
\caption{ Charge asymmetry quantities and their statistical
uncertainties for the process
$\gamma_{\lambda_1}\gamma_{\lambda_2}\to W\mu\nu$.} \label{tABI}
\end{table*}
\end{center}
We start with the study of asymmetry neglecting  the
cascade channel and supposing photon beams monochromatic and
completely polarized. First, we present  the distributions
$\partial^2 \sigma/(\partial p_{\parallel}\partial p_{\perp})$ of
muons in the ($p_\|,\,p_\bot$) plane, at different photon
polarizations, in Figs.~\ref{asymm_no_tau+-}. These figures show
explicitly strong difference in the distributions of negative and
positive muons as well as strong dependence of distributions on
photon polarizations. Therefore, the charge asymmetry in the process
is {\it a strong effect}.
\begin{figure}[t]
\includegraphics[scale=0.5]{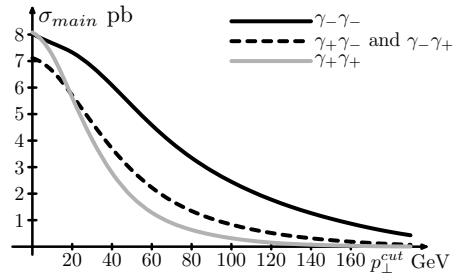}
\caption{Dependence of cross sections on cut $p_{\bot\mu}^c$ for
main process.  } \label{figsigmas}
\end{figure}
\begin{figure}[b]
\includegraphics[scale=0.5]{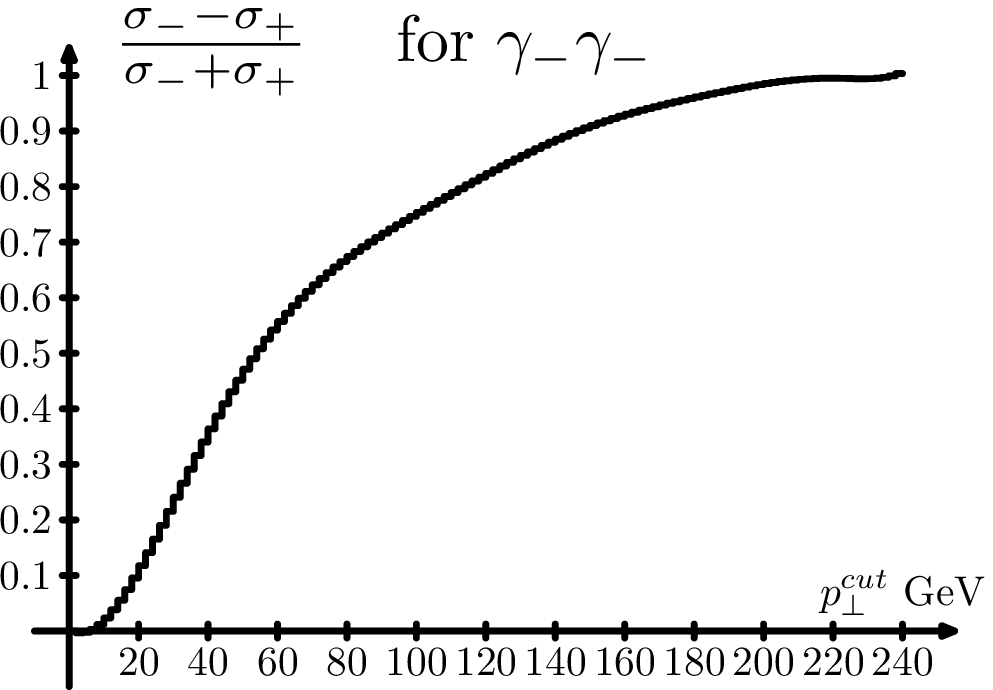}
\includegraphics[scale=0.5]{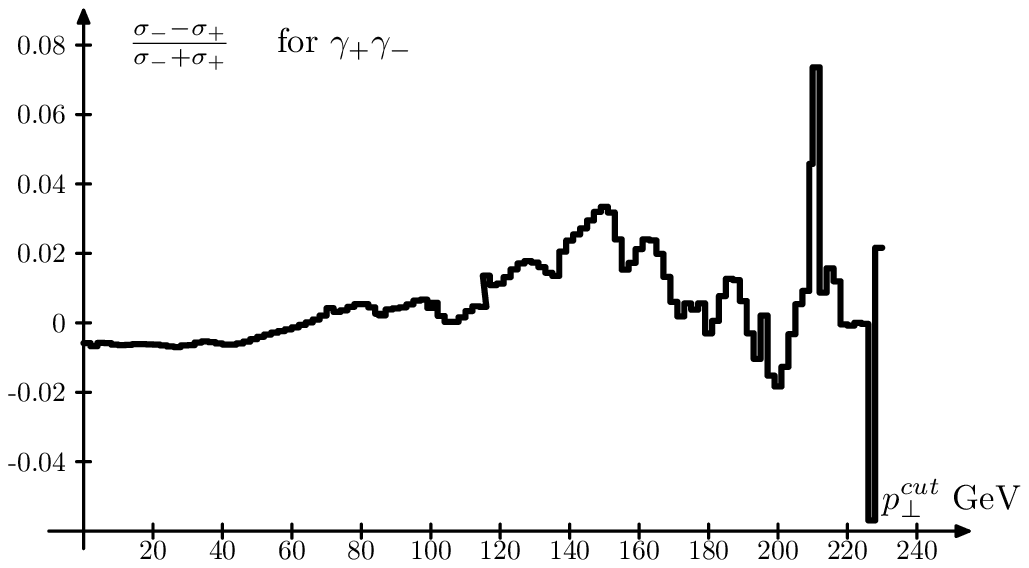}
\caption{Relative cross section differences in dependence on cut
$p_{\bot\mu}^c$ value.} \label{figcrsecasyn}
\end{figure}

Table~I presents obtained average momenta for the negative and
positive muons and corresponding asymmetry quantities~\eqref{eqPLT}
together with their statistical uncertainties (in percents)  for
different cuts $p_{\bot\mu}^c$. One can see that the values of
asymmetry are typically 20-50\%.

We have checked that with the change of sign of both photon
helicities mean muon momenta for negative and positive muons change
their places (within statistical accuracy) so that the quantities
$\Delta_{L,T}$ change their signs with this change of polarization.

The longitudinal scale of distributions in momenta is determined by
initial photon energy while the transverse scale is determined by
the $W$ mass. Hereupon mean transverse momenta are usually smaller
than longitudinal.

Besides, for the collision of photons with identical helicity at the
growth of cut $p_{\bot\mu}^c$ the cross sections for production of
positive and negative muons become different, due to discussed
charge asymmetry (remind, we discuss only events with negative
muons or $W$'s flying in forward hemisphere). For the collision of
photons with opposite helicity these cross sections should coincide
since, for example, for $\gamma_-\gamma_+$ collision forward
hemisphere for $\mu^-$ production realize absolutely the same
distribution as backward hemisphere for $\mu^+$ production. This
effect is clearly seen from Figure~\ref{figsigmas}, where we present
the $p_{\bot\mu}^c$ dependence for  cross sections \eqref{relcrsec}
at different initial photon polarizations.

The corresponding cross section differences, Eq.~\eqref{crsecdif},
are shown in Figure~\ref{figcrsecasyn} {\it without averaging over
realizations}. The difference in cross sections for
$\gamma_-\gamma_-$ and $\gamma_+\gamma_+$ collisions is due to our
choice of events with negative particles in forward hemisphere. In
this case the cross sections coincide at small $p_{\bot\mu}^c$,
while at high $p_{\bot\mu}^c$ one of them becomes larger and larger
in comparison with the other, thus $\Delta\sigma$ goes to one (one
more demonstration of transverse asymmetry). For the case with
opposite photon helicities the deviation of $\Delta\sigma$ from zero
shows high statistical uncertainty, due to the very low value of the
cross sections (low counting rates) at $p_{\bot\mu}>120$~GeV.
Similar dependencies with roughly the same characteristic values of
$p_{\bot\mu}^c$ remain valid even at higher collision energies (for
example, at $\sqrt{s}=2$~TeV) since the scale of this dependence is
determined by the $W$ boson mass and not by  the total energy.

\section{Global asymmetries in cascade process with intermediate {\boldmath $\tau$}}
\label{seccasccontr}

The observable final state with two muons or $W+\mu$ with missing
transverse momentum carried away by neutrinos can appear either via
processes  $\gamma\gamma \to \mu^+\mu^- \nu_{\mu} \bar{\nu}_\mu$
($\gamma\gamma \to W\mu \nu$) or via cascade processes with $\tau$
production  and subsequent $\tau$ decay ($\tau \to
\mu\nu_\mu\nu_\tau$), see  Eq.\eqref{cascadeproc}. The latter
process enhances the total event rate (without cuts) by a value
given by factor $B\equiv Br(\tau\to\mu\nu\nu)=17\,\%$ for the
$\ggam\to W\mu+\nu's$. Similar event rate enhancement in the process
$\ggam\to\mu^+\mu^-+\nu's$ is $2B+B^2\approx 37\,\%$.

The accurate calculation of processes with six or more particles in
the final state like\lb $\gamma\gamma \to \mu^+\mu^- \nu_{\mu}
\bar{\nu}_\mu\nu_{\tau} \bar{\nu}_\tau$ is a computationally
challenging task with available software. Since $\tau$ is very
narrow particle, the diagrams without $\tau$-pole in $s$-channel can
be neglected with very high precision, ($\sim \Gamma_\tau/m_\tau$).
Therefore, one can in principle use the results for $\ggam\to
W\tau\nu$ ($\gamma\gamma \to \tau^+\mu^- \nu_{\mu} \bar{\nu}_\mu$,
etc.) and convolute them with the distribution of $\mu$ from $\tau$
decay. However, the latter distribution depends on $\tau$
polarization which cannot be determined definitely with
CompHEP/CalcHEP in general case. (Generally, all helicity amplitudes
for $\tau$ production are nonzero, and for convolution one must
consider not only diagonal helicity states but also their
interference).

Fortunately, the cascade process provides only a small fraction of
the total cross section and the main contribution to the total cross
section is given by the double resonant  (DRD) diagrams of
Fig.~\ref{figdiagwmu}(a). That is the reason why in the  description
of the cascade contribution only these diagrams can be taken into
account --- {\it DRD approximation}. In this approximation the
$\tau$ helicity is precisely determined in each MC event. In
sect.~\ref{secinaccest} we will show that the inaccuracy introduced
in the total result using the  DRD approximation for the cascade
contribution is within the estimated statistical uncertainty, found
for the main process.

Note that the distributions obtained in sect.~\ref{secasWmu}
describe with high accuracy also $\tau$-distributions  in $\ggam\to
W\tau\nu$ processes etc.

In the DRD approximation each $\tau$ is produced only via $W$-decay,
and its  polarization  in the rest frame of $W$ is given by the SM
vertex, $\tau^+ W^-_\mu\gamma^\mu(1-\gamma^5)\nu_\tau +h.c.$. Due to
$\gamma^\mu(1-\gamma^5)$ factor, $\tau$ helicity is opposite to that
of $\nu_\tau$, it is positive for $\tau^+$ and negative for $\tau^-$
(with accuracy to $m_\tau/M_W$) and independent on $W$ polarization.

For each generated event momenta of all particles are known. The
spin vector of $\tau$ is expressed easily via momenta of $\tau$ and
$\nu_\tau $, $p_\tau$ and $p_\nu$ respectively, as
 \be
\pm s/2,\;\mbox{ where }\; s=\left(\fr{p_\tau}{m_\tau}-\fr{p_\nu
m_\tau}{(p_\tau p_\nu)} \right)\qquad
\left\{\begin{array}{c} +\;\;\mbox{ for } \tau^+,\\
-\;\;\mbox{ for } \tau^-.\end{array}\right.
 \label{eqspin}
 \ee

Denoting the momentum of $\mu$ by $k$, the distribution of muons in
$\tau$-decay with momentum $p_\tau$ and spin $\pm s$ can be written,
neglecting muon mass, as
 \be
f=\dfrac{4}{\pi E_\tau m_\tau^5}\left[(3m_\tau^2-4p_\tau k)p_\tau k+
k s\cdot m_\tau(4p_\tau k-m_\tau^2)\right]d\Gamma,
\,\label{eqmudistr}
 \ee

where $d\Gamma$ is a phase space element boosted to the lab frame.
In the $\tau$ rest frame $d\Gamma =
\theta\left(m_\tau/2-k\right)d^3k/E_\mu$. Note that the sign of
helicity before $s$ in Eq.~\eqref{eqspin} disappears in the result.

\begin{figure}[b]
\begin{center}
\includegraphics[scale=0.23]{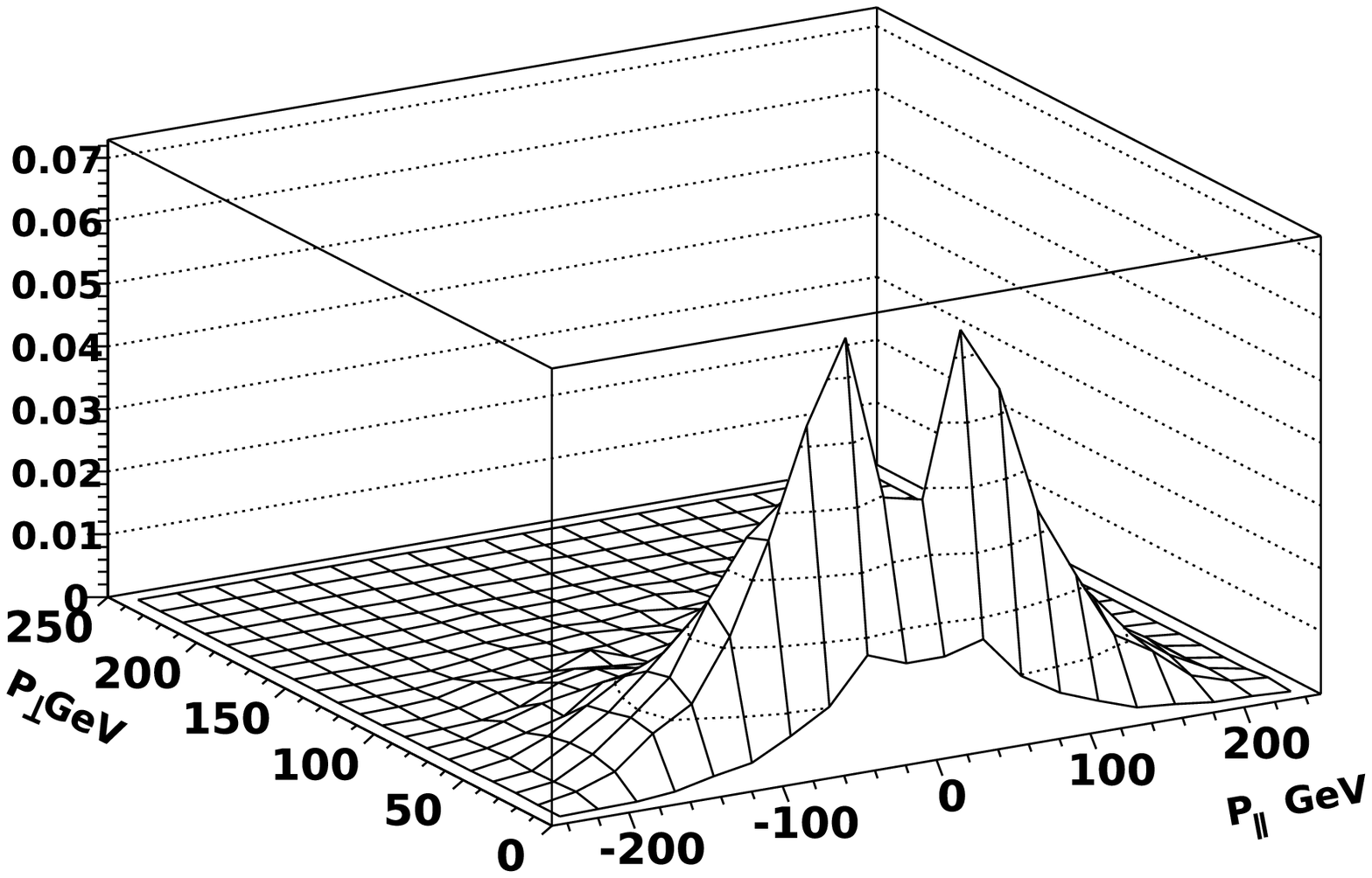}
\includegraphics[scale=0.23]{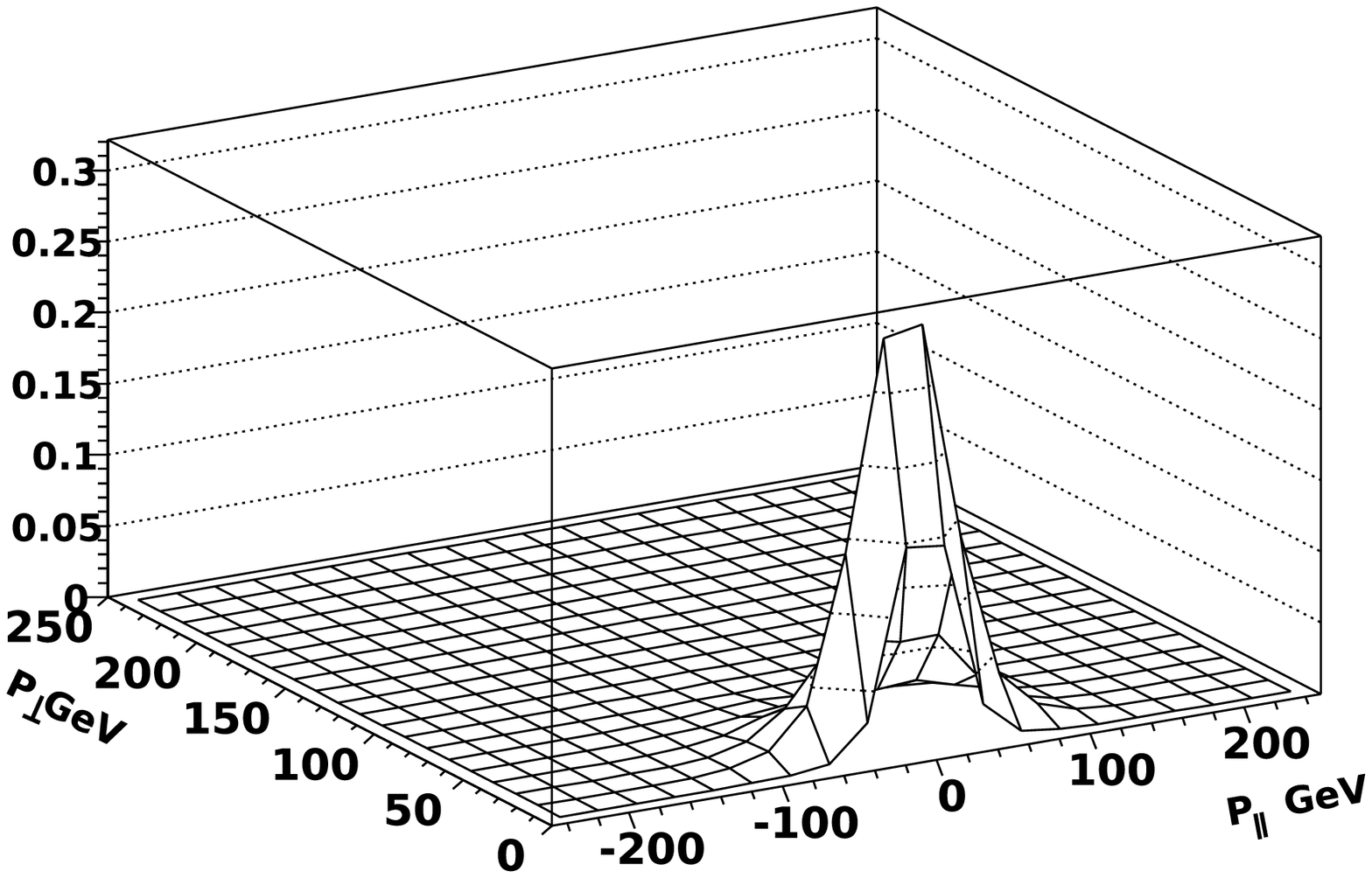}\\
\includegraphics[scale=0.23]{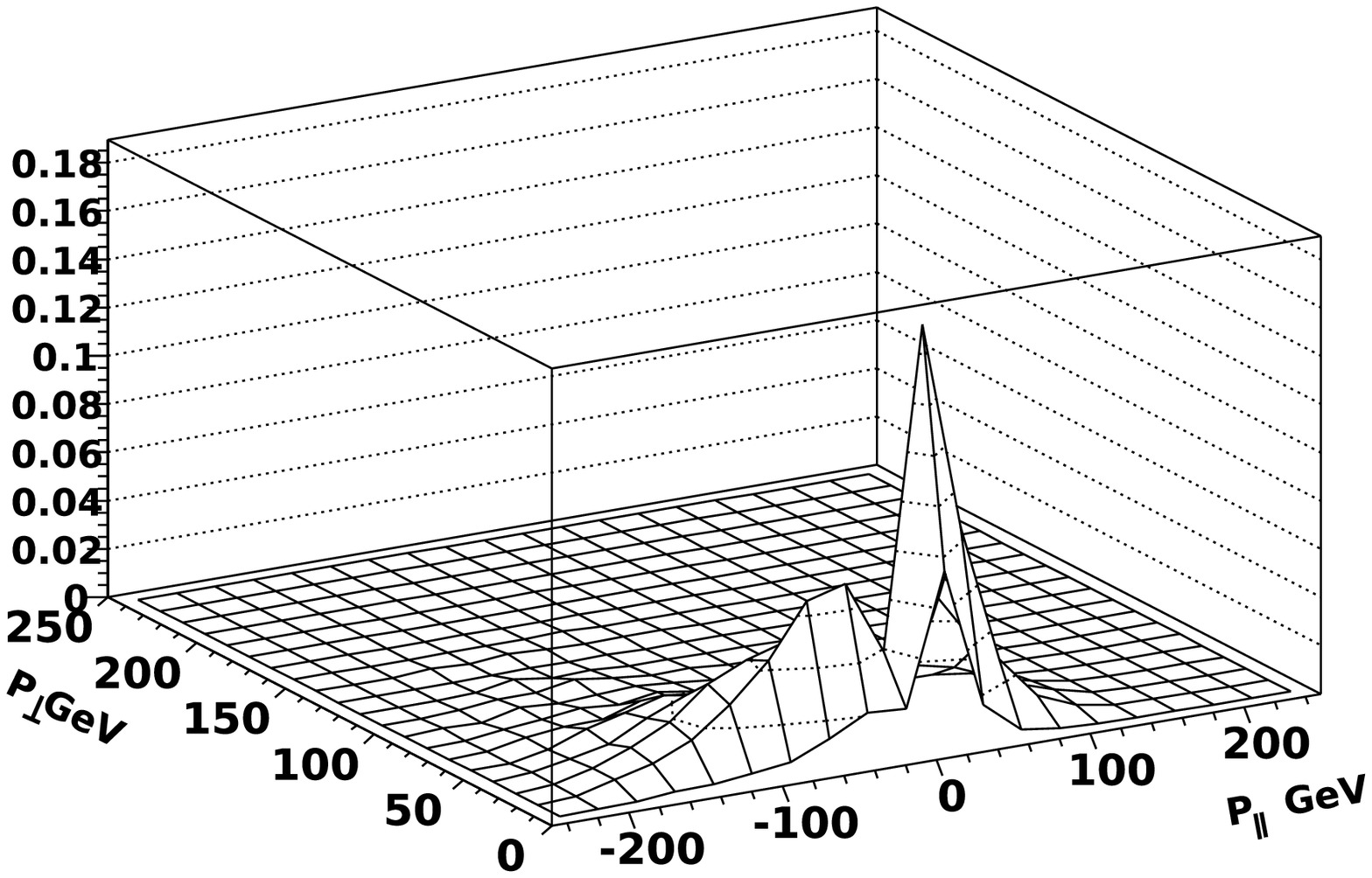}
\includegraphics[scale=0.23]{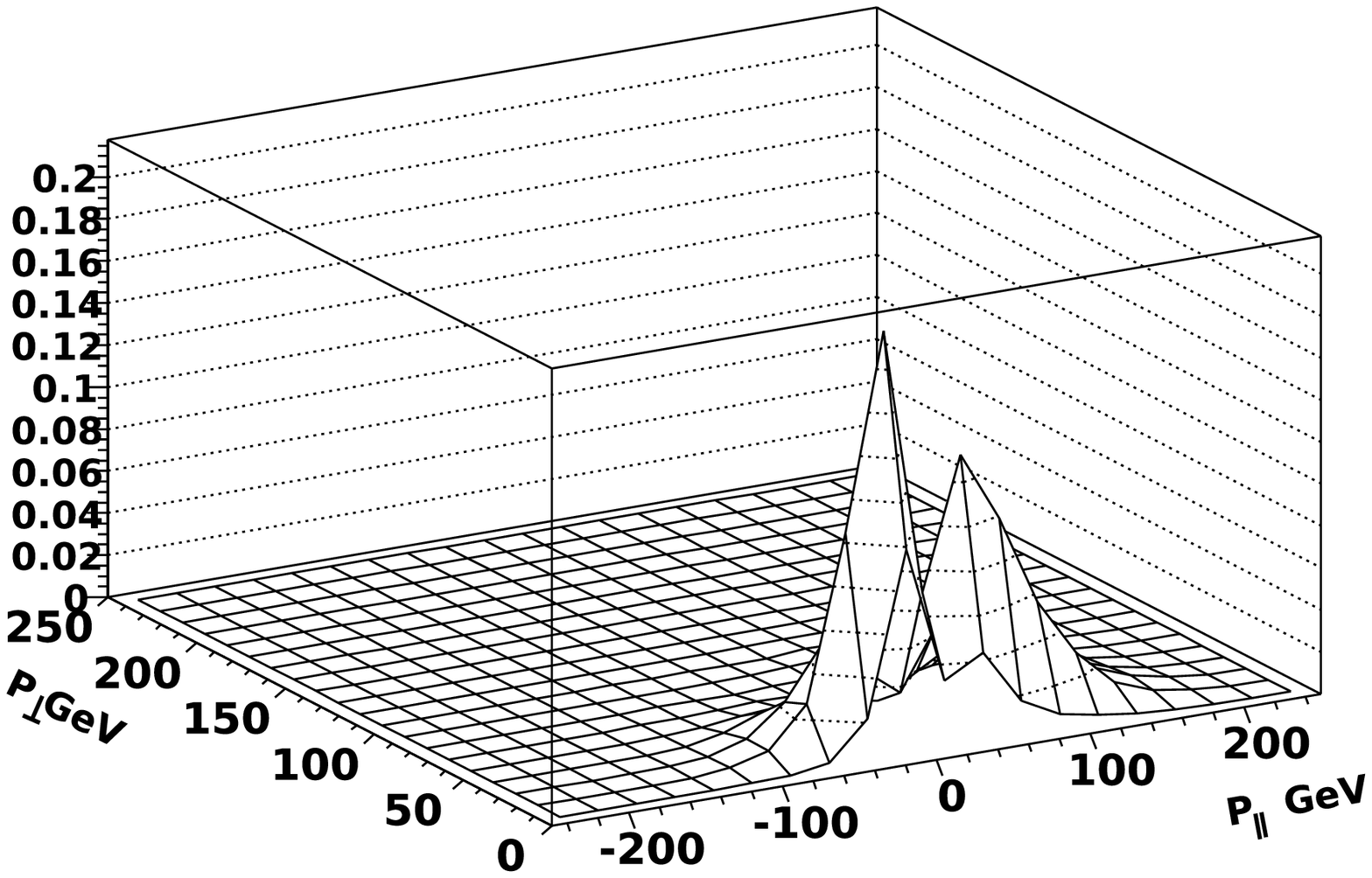}
\caption{Muon distribution  in cascade processes
$\gamma_-\gamma_-\to W\mu\nu\nu\nu$ (upper plot) and
$\gamma_+\gamma_-\to W\mu\nu\nu\nu$ (lower plot), left --
$\mu^-$,right -- $\mu^+$}
\end{center}
\label{asymm_From_tau+-}
\end{figure}

Let us discuss now the qualitative  features of the muon spectrum
given by the convolution of the $\tau$ spectrum with  the
distribution in Eq.~\eqref{eqmudistr}. One can consider
$\tau\to\mu\nu\nu$ decay as a two body decay: a massless muon and
the di-neutrino with invariant mass $m_{\nu\nu}$. At given
$m_{\nu\nu}$, the energy and 3-momentum of the muon in the $\tau$
rest frame are $\vep_\mu = p_\mu^0= Am_\tau/2$ with
$A=1-(m_{\nu\nu}/m_\tau)^2$. In the laboratory frame where the
3-momentum of the $\tau$,  $\vec{p}_{\tau}$ is under some angle
$\theta$ relative to muon momentum in $\tau$ rest frame, the muon
momentum is evidently $\vec{p}_{\mu}=A\vec{p}_\tau (1+\cos\theta)/2$
plus small corrections negligible in our discussion. Therefore, the
muon distribution repeats in some sense that of the $\tau$ but with
a factor $A(1+\cos\theta)/2$, which is usually much lower than one.
In other words, the distribution of muons in the cascade process is
similar in the main features to that of the $\tau$  but it is {\it
strongly contracted to the origin of the coordinates}.

\bu \ It is useful to describe  the  inaccuracy of the DRD
approximation  in the description of the $\ggam\to W\tau\nu$ cross
section itself, $\delta^{\tau W}_{DRD}$, in dependence on the cut
$p_{\bot\tau}^c$. The estimates in Sec.~\ref{secdiagr} show that at
the considered energies all contributions to the cross section are
small in comparison with that of the DRD except for the SRDW
contribution. The interference term $Re(A_{DRD}^*A_{SRDW})$ is
roughly of the same order of magnitude as $|A_{SRDW}|^2$ since the
DRD is large only in regions of the final phase space corresponding
to the $W$ resonances, while the other contributions do not have
these peaks. The numerical value of this inaccuracy is obtained by
direct comparison of this cross section, calculated with all
diagrams, and those for DRD diagrams with MC simulation.

At $p_{\bot\tau}^c=10$~GeV we find that for the $\ggam\to W\tau\nu$
process the SRDW contribution itself is about 5~\% of the DRD one,
and the interference of this contribution with DRD is destructive so
that the DRD contribution differs from total cross section only by
about 1\%. This difference naturally grows with increasing values of
$p_{\bot\tau}^c$.

More important for us is the inaccuracy of DRD approximation in the
description of asymmetry quantities \eqref{eqPLT}.
Table~\ref{Tabinac} presents value of inaccuracy $\delta^{\tau
W}_{DRD}(p_{\bot\tau}^c)$ introduced by DRD approximation in the
description of  $\ggam\to W\tau\nu$ process at different cut values
of {\it $\tau$ transverse momenta $p_{\bot\tau}^c$}.
\begin{table}[t]
\begin{tabular}{||c||c|c||c|c||}\hline\hline
{$p_{\bot \tau}^c$, (GeV)}& $\delta_L^{(--)} (\%) \, $ & $\delta_T^{(--)}
(\%) \, $ & $\delta_L^{(+-)} (\%) \, $ & $\delta_T^{(+-)} (\%) \, $\cr
 \hline
 10 & 0.9&2.3& 0.7&3.45\cr
 40& 1.5&3.6&2.1&4.1\cr
 80& 1.9&5.6&4.2&7.7\cr
 120 & 5.7& 5.3& 4.4&31\cr\hline\hline
\end{tabular}
\caption{ Inaccuracy of DRD approximation $\delta^{\tau
W}_{DRD}(p_{\bot\tau}^c)$ for $\Delta_{L,T}$ at different $p_{\bot
\tau}^c$ for  $\tau$ production.}
\label{Tabinac}
\end{table}

Large values of the relative quantity $\delta_T^{(+-)}$ at large
$p_{\bot \tau}^c$ appear in the case when the absolute value of
$\Delta_T^{(+-)}$ is negligibly small. One can see that this
inaccuracy grows with increasing the value of the  cut, see
Fig.~\ref{figcrsecasyn}. However with this increase also the
fraction of cascade process within the total process becomes smaller
and smaller (see discussion at the end of next section).

For the processes $\ggam\to \mu^\pm\tau^\mp\nu\nu$ and $\ggam\to
\tau^+\tau^-\nu\nu$ the inaccuracies of DRD approximation are
$\delta^{\tau \mu}_{DRD}(p_{\bot\tau}^c)= \delta^{\tau
W}_{DRD}(p_{\bot\tau}^c)$ and $\delta^{\tau
\tau}_{DRD}(p_{\bot\tau}^c)= 2\delta^{\tau
W}_{DRD}(p_{\bot\tau}^c)$.

\section{Total asymmetries} \label{secresult}

The resulting distributions include the complete tree-level results
of $\ggam\to W\mu\nu$ and DRD approximation for cascade
contribution.
\begin{figure}[t]
    \begin{center}
 \includegraphics[scale=0.23]{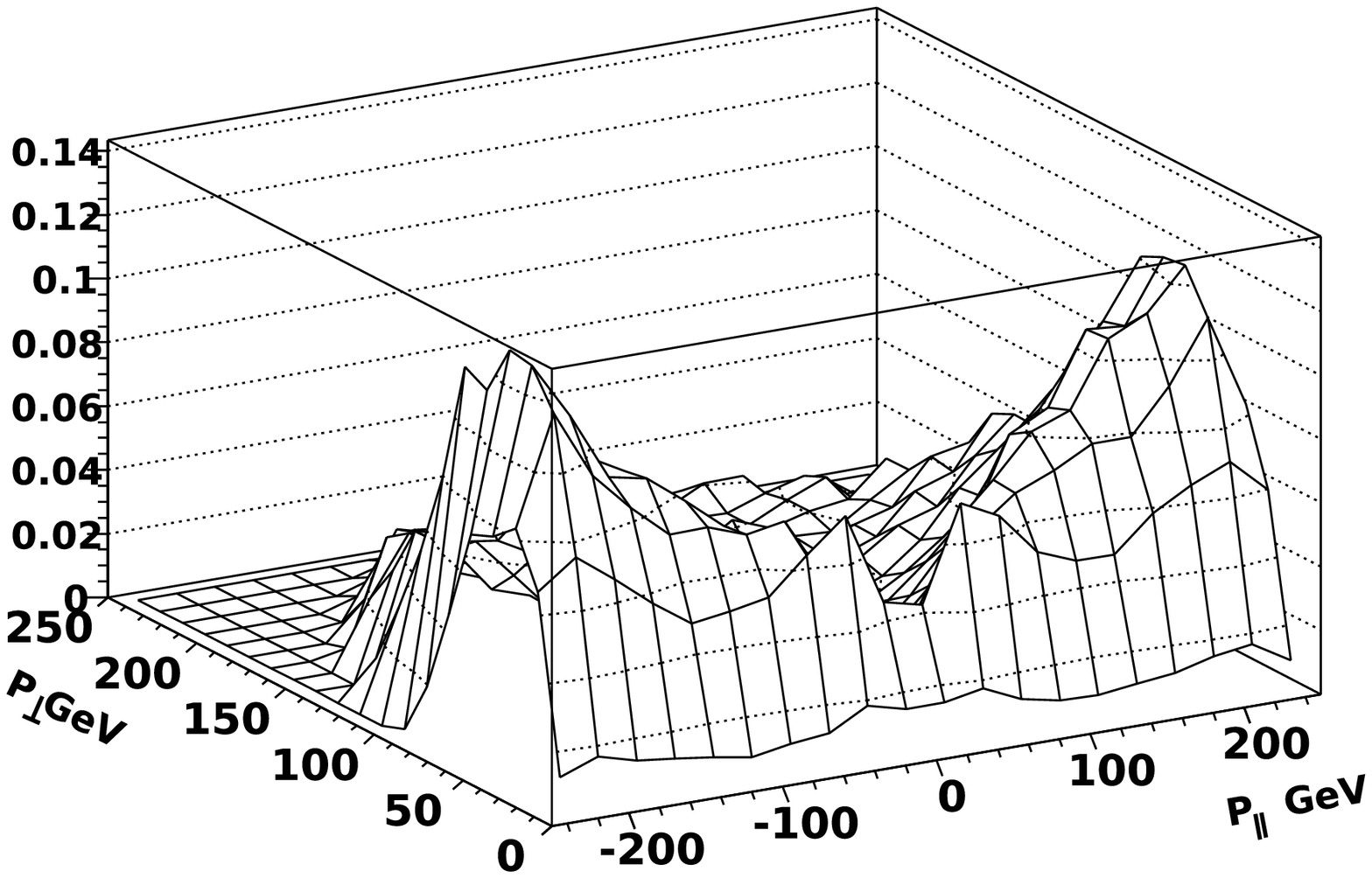}
 \includegraphics[scale=0.23]{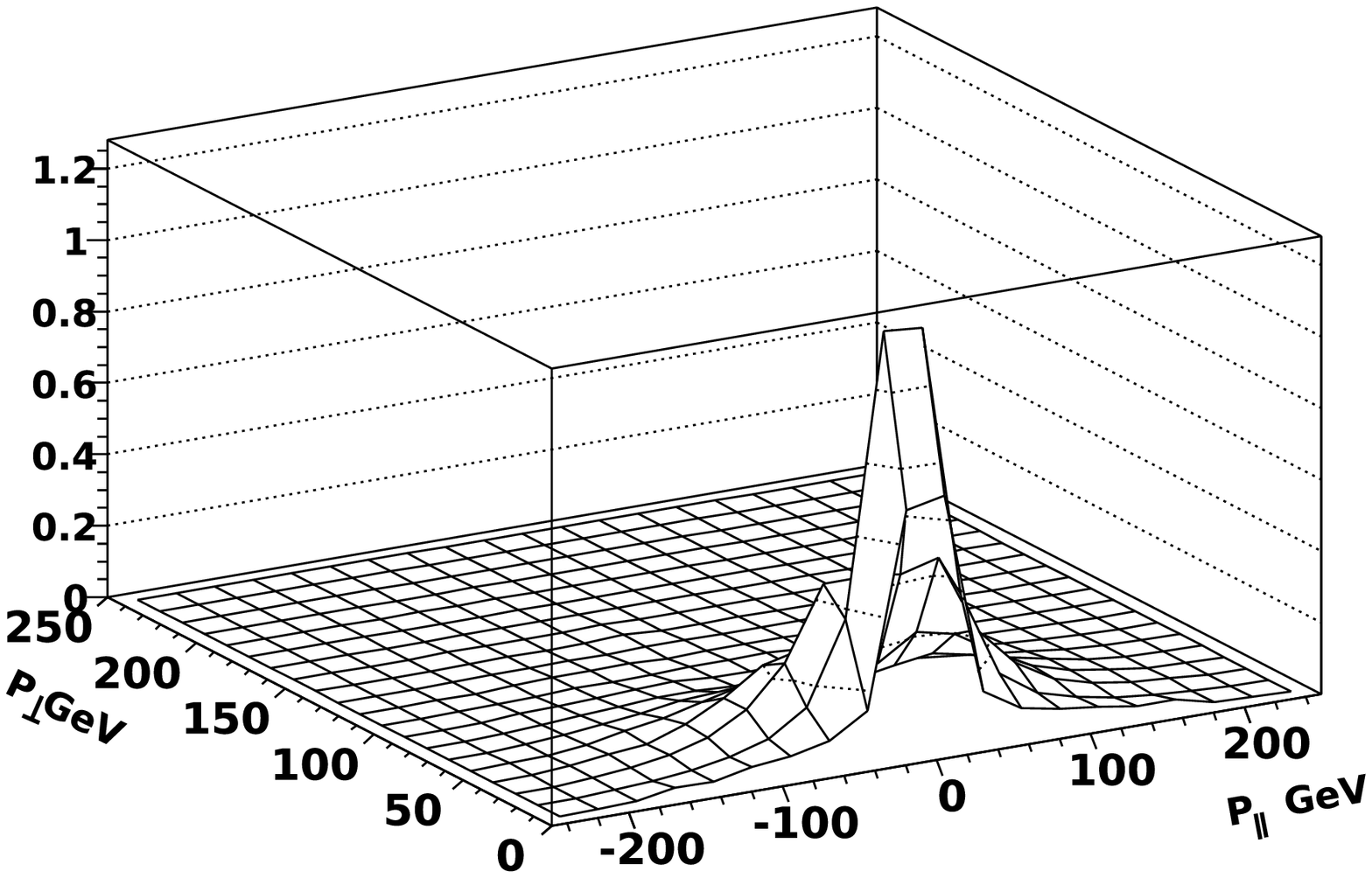}\\
 \includegraphics[scale=0.23]{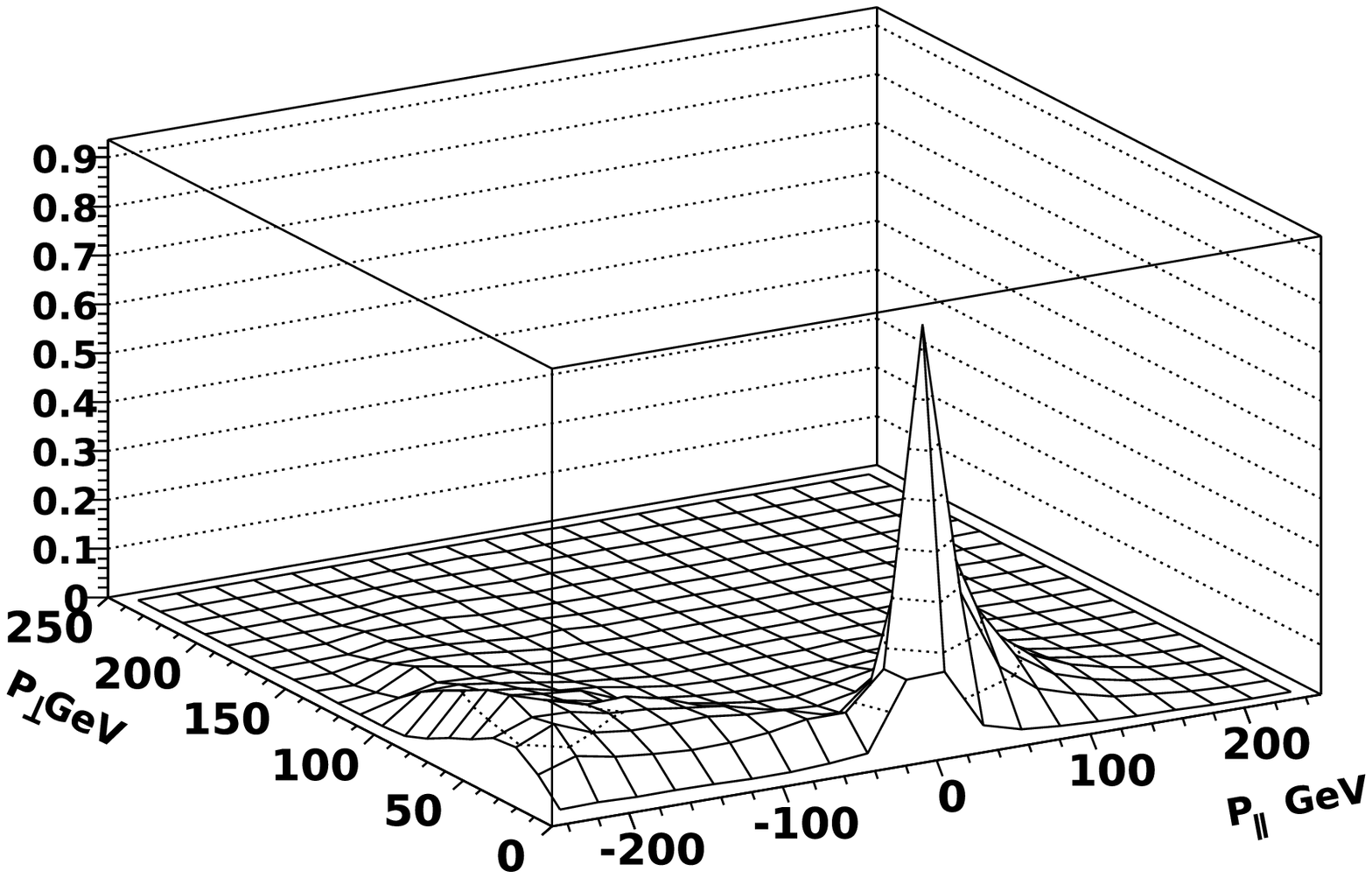}
 \includegraphics[scale=0.23]{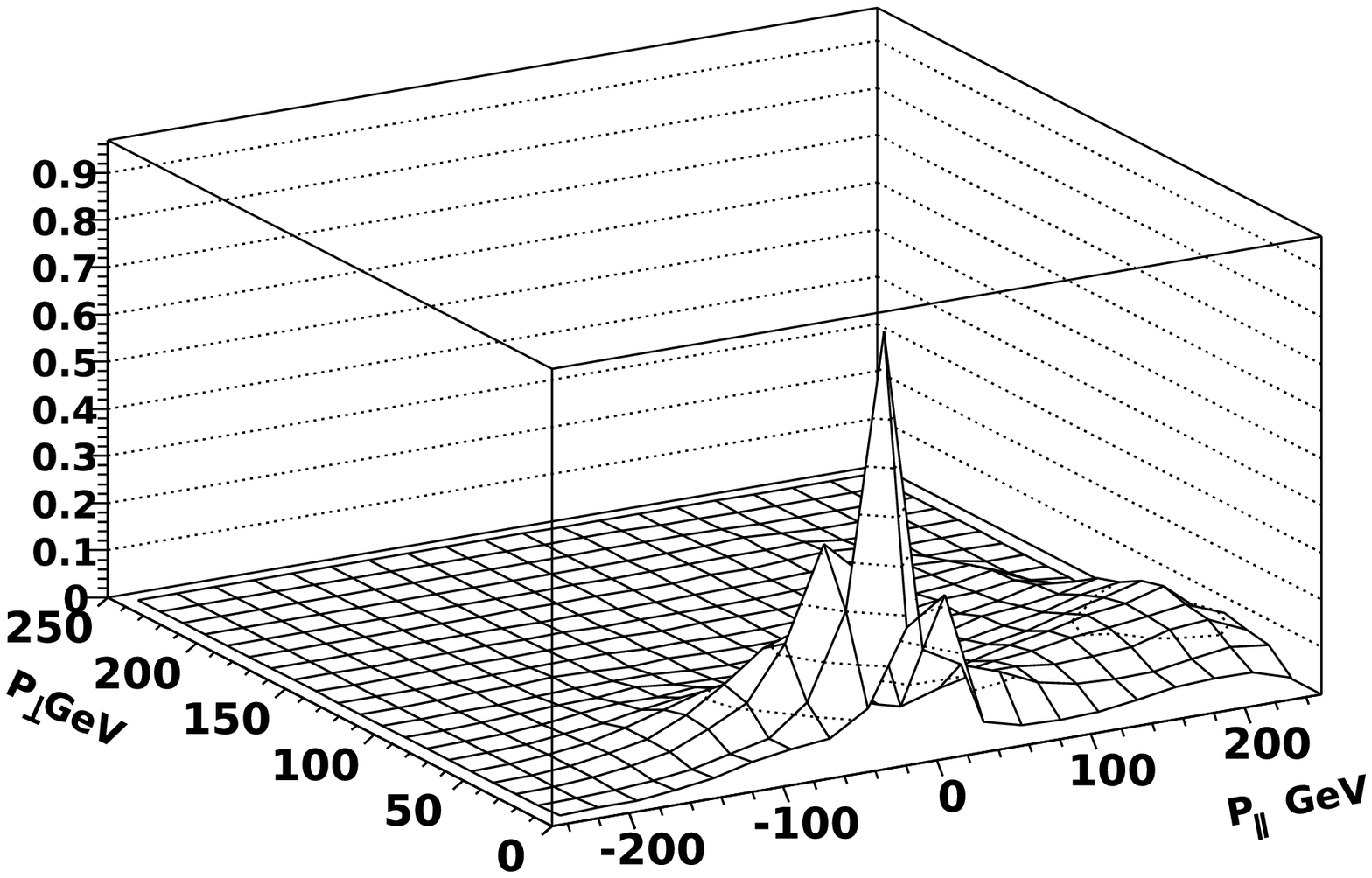}
\caption{Total muon distribution in
$\gamma_-\gamma_-\to W\mu+\nu's$ (upper plot) and
        $\gamma_+\gamma_-\to W\mu+\nu's$ (lower plot);  left
-- $\mu^-$,right -- $\mu^+$} 
        \label{asymm_With_tau+-}
    \end{center}
\end{figure}

Figures~\ref{asymm_With_tau+-} show the total observable
distributions of muons, i.e. the sum of distributions of muons in
$\ggam\to W\mu\nu$ and $\ggam\to W\tau\nu\to W\mu\nu\nu\nu$, and
Table~\ref{TabresultW} presents the corresponding total asymmetry
quantities for $p_{\bot \mu}^c=10$ GeV.
\begin{table}[b]
\begin{tabular}{||c|c|c|c|c|c|c||}\hline\hline
$\, \gamma_{\lambda_1}\gamma_{\lambda_2} \,$ & $\,P_L^- \, $&
$\,P_L^+ \, $& $\,\Delta_L \, $&$\,P_T^- \, $& $\,P_T^+ \, $&
$\,\Delta_T \, $\cr\hline $\, \gamma_{-}\gamma_{-} \,$ & $\, 0.548\,
$& $\,0.164\, $& $\, +0.539 \, $&$\,0.311 \, $& $\,0.142 \, $& $\,
+0.374 \, $\cr\hline $\, \gamma_{+}\gamma_{-} \,$ & $\, 0.199\, $&
$\,0.513\, $& $\, -0.440 \, $&$\,0.152 \, $& $\,0.232 \, $& $\,
-0.207 \, $\cr\hline\hline
\end{tabular}
\caption{ Resulting asymmetry quantities.}
\label{TabresultW}
\end{table}
Comparison with Fig.~\ref{asymm_no_tau+-} and Table~I shows that the
cascade process introduces a change in the shape of muons
distribution only at small momenta and its contribution reduces the
asymmetry parameters $\Delta_{L,T}$ in average by about 3~\% only.

\begin{figure}[hbt]
\begin{center}
\includegraphics[scale=0.45]{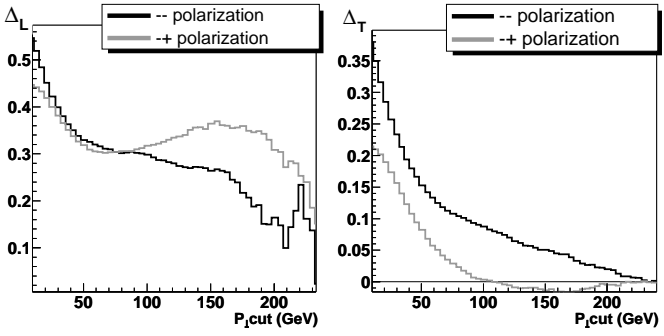}
\caption{ The $p_{\bot \mu}^c$ dependence of asymmetry.
Left plot -- $\Delta_L$, right plot -- $\Delta_T$, black
lines -- for $\gamma_-\gamma_-$, gray lines -- for
$\gamma_-\gamma_+$} \label{cut_dependence_as}
\end{center}
\end{figure}
The plots in Fig.~\ref{cut_dependence_as} show the dependence of
asymmetries $\Delta_L$ and $\Delta_T$ on $p_{\bot \mu}^c$. The
longitudinal charge asymmetry remains large even with large cuts,
while the transverse charge asymmetry diminishes with $p_{\bot
\mu}^c$ growth. In particular, for $\gamma_+\gamma_-$ collision at
$p_{\bot \mu}^c \geq 120$~GeV the quantities $P_T^+$ and $P_T^-$
practically coincide, giving negligible $\Delta_T$, right plot, with
naturally high statistical uncertainty in this small quantity.

\begin{figure}[b]
\includegraphics[scale=0.6]{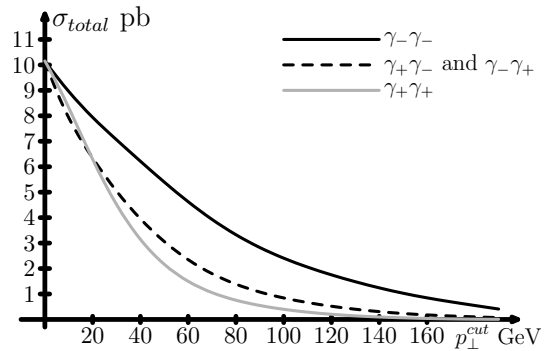}
\caption{
The $p_{\bot \mu}^c$ dependence of resulting
cross sections for different photon polarizations.}
   \label{cut_dependence_fullcs}
 \end{figure}

The $p_{\bot \mu}^c$ dependence of resulting cross sections for
different photon polarizations is shown in
Fig.~\ref{cut_dependence_fullcs}. The difference in curves for
$(++)$ and $(--)$ initial states arise because of  our {\it charge
asymmetric} selection of  events,  with negative particles flying in
the forward hemisphere (see Sec.~\ref{secglob}).

\subsection{Inaccuracy of DRD approximation for
resulting asymmetries}\label{secinaccest}

Let us denote by $\delta^{tot}_{DRD}(p_{\bot \mu}^c)$ the inaccuracy
of the DRD approximation for the resulting asymmetries, by
$\delta^{casc}(p_{\bot \mu}^c)$ --- the inaccuracy of DRD
approximation for the description of the cascade process itself,
like that given in the Table~\ref{Tabinac}, and by
 \begin{equation}
R(p_{\bot \mu}^c)=\frac{\sigma^{casc}(p_{\bot \mu}^c)}{\sigma^{tot}(p_{\bot \mu}^c)}
\label{relativcontrib}
 \end{equation}
the relative contribution of cascade $\mu$ in the total cross
section, all -- in dependence on cut for muons $p_{\bot \mu}^c$.
Naturally,
 \be
\delta^{tot}_{DRD}(p_{\bot \mu}^c)=
R(p_{\bot \mu}^c)\, \delta^{casc}(p_{\bot \mu}^c)\,. \label{inaccest}
 \ee
\begin{figure}[hbt]
\includegraphics[scale=0.6]{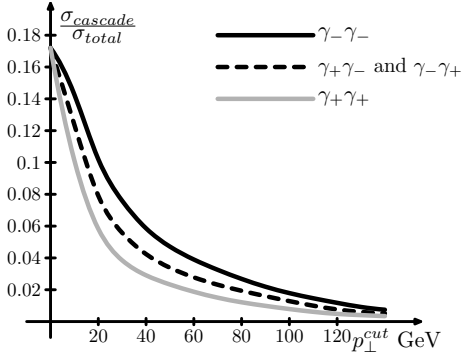}
\caption{
Relative contribution of the cascade process $R(p_{\bot \mu}^c)$ as defined in Eq.\protect\ref{relativcontrib} vs
$p_{\bot\mu}^c$.  }
   \label{figtaucontr}
 \end{figure}

Due to the contraction of the distribution of muons produced in the
cascade process in comparison with that of the parental $\tau$ (see
Sec.\ref{seccasccontr}), and consequently with that of $\mu$ in the
main process, the relative contribution of the cascade $\mu$ in the
total cross section $R(p_{\bot \mu}^c)$ falls rapidly with the
growth of the cut $p_{\bot\mu}^c$, as can be seen in
Fig.~\ref{figtaucontr}.

At $p_{\bot\mu}^c=0$ we have $\delta^{casc}(p_{\bot
\mu}^c)=\delta^{casc}(p_{\bot \tau}^c)$. With the numbers given by
Table~\ref{Tabinac} and Fig.~\ref{figtaucontr} one can see that the
inaccuracies \eqref{inaccest} are lower than the expected
statistical uncertainty of future experiments (Table~\ref{tABI},
first two lines). With growth of $p_{\bot\mu}^c$ the inaccuracy
$\delta^{casc}(p_{\bot \mu}^c)$, similar to that given in the
Table~\ref{Tabinac}, increases, but the  cascade contribution
$R(p_{\bot \mu}^c)$, Fig.~\ref{figtaucontr}, decreases faster.
Therefore the resulting inaccuracy introduced by the DRD
approximation in Eq.\eqref{inaccest} is well within the expected
statistical uncertainty of future experiments, Table~\ref{tABI},  at
each cut on transverse momentum for the $\ggam\to W\mu+\nu's$
process (and it is within the expected statistical uncertainty of
future experiments for $\ggam\to\mu^+\mu^-+\nu's$ process).

\section{Effect of photon non--monochromaticity}
\label{non-mon}

At the PC photons will be non-monochromatic with spectra peaked near
the high energy limit $E_\gamma^{max}$. Moreover, due to the finite
distance between the conversion point (CP) and the interaction point
(IP) and also due to rescatterings of laser photons on electrons
after the first collision, photon spectra are even non-factorizable.
Fortunately in their high energy part,
($E_\gamma>E_\gamma^{max}/\sqrt{2}$), these spectra are factorizable
with a high precision and these photons have a high degree of
polarization. Moreover, the form of the effective spectra in this
region is described with high accuracy with the aid of only one
additional parameter, independent from the  details of the
experimental setup, while the polarization is the same as for pure
Compton effect~\cite{GinK}. {\it The luminosity of the Photon
Collider is normalized for this very region only.}

\begin{figure}[t]
\includegraphics*[scale=0.6]{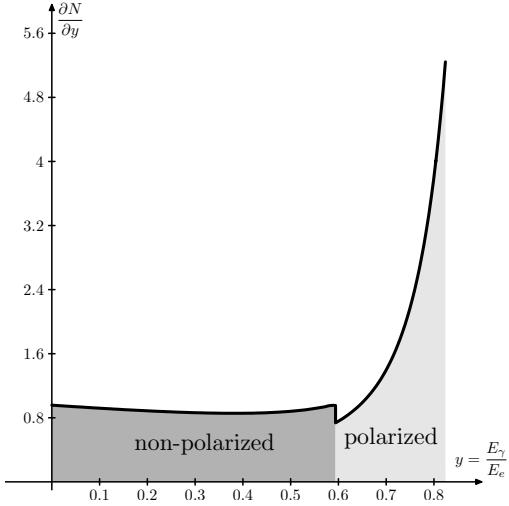}
\caption{The "realistic" photon spectra, used in our calculations.}
\label{photspec}
\end{figure}
The low energy part of the effective photon spectrum,  depends
strongly on the details of the experimental setup which may change
during the construction process of  the ILC. Therefore, in our
simulations, we used a photon spectrum composed of two parts as
shown in Fig.~\ref{photspec}. At $E_\gamma>E_\gamma^{max}/\sqrt{2}$
we used the approximation from Ref.~\cite{GinK} with $\rho=1$ and
$x=4.8$ with polarization for ideal Compton effect~\cite{Ginzburg1}.
In order to imitate the low energy part of the spectrum (at
$E_\gamma<E_\gamma^{max}/\sqrt{2}$) we used spectra
from~\cite{Ginzburg1} for the case when the IP and CP coincide
($\rho=0$) and consider these photons to be unpolarized.

In this section we denote by $\gamma_-$ an initial photon state
obtained in collision of laser photon with helicity $P_c=+1$ and an
electron with mean double initial helicity $2\lambda_e=0.85$, which
gives $\lambda_\gamma=-1$ for photons with
$E_\gamma=E_\gamma^{max}$. In this case mean polarization of photons
with $E_\gamma>E_\gamma^{max}/\sqrt{2}$ is also negative but its
absolute value is lower than one as it is described in
\cite{Ginzburg1}. At $E_\gamma<E_\gamma^{max}/\sqrt{2}$ we treat
these photons as non-polarized. The state $\gamma_+$ is defined in
the same way.

\begin{figure}[b]
    \begin{center}
 \includegraphics[scale=0.23]{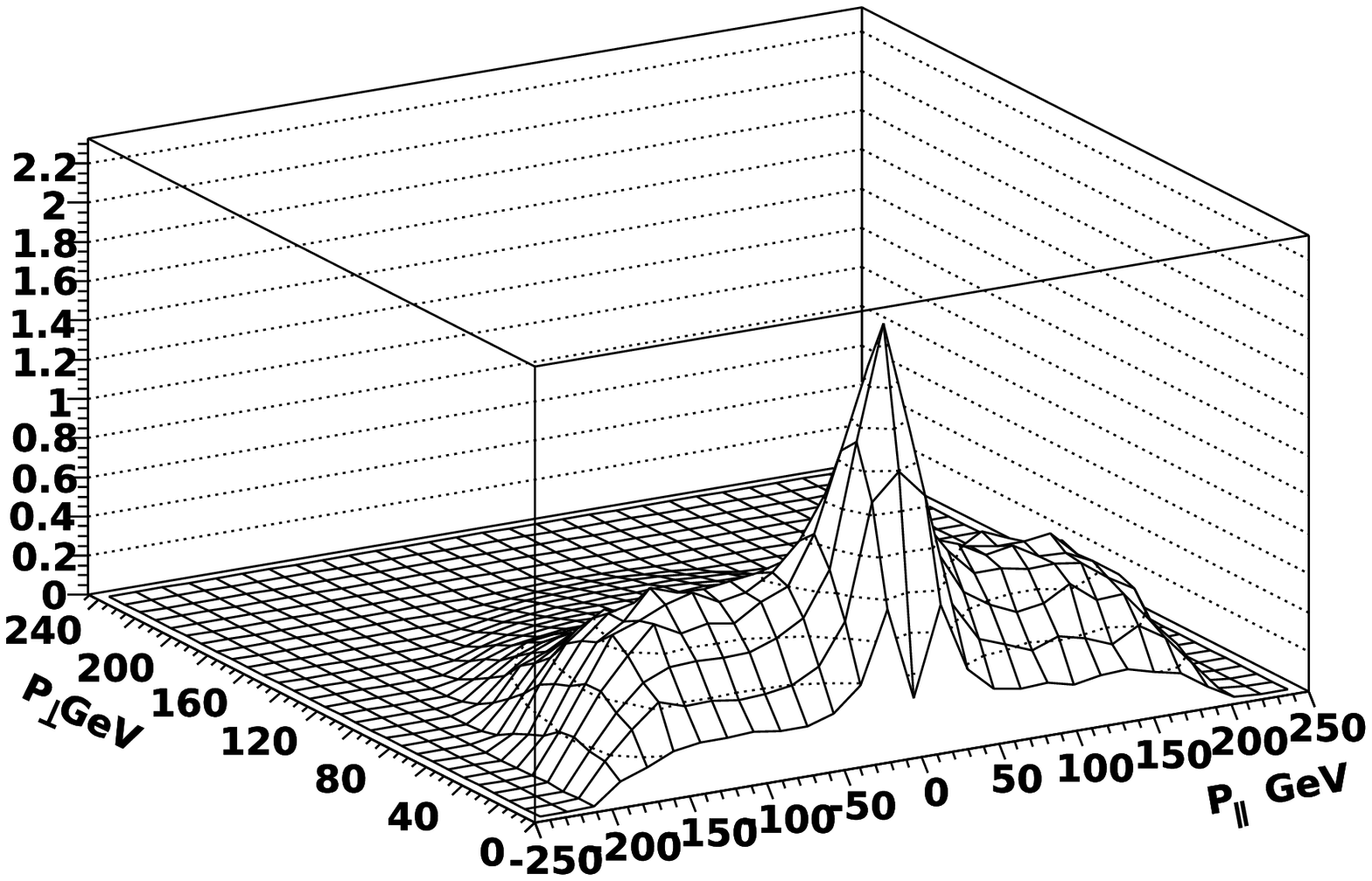}
 \includegraphics[scale=0.23]{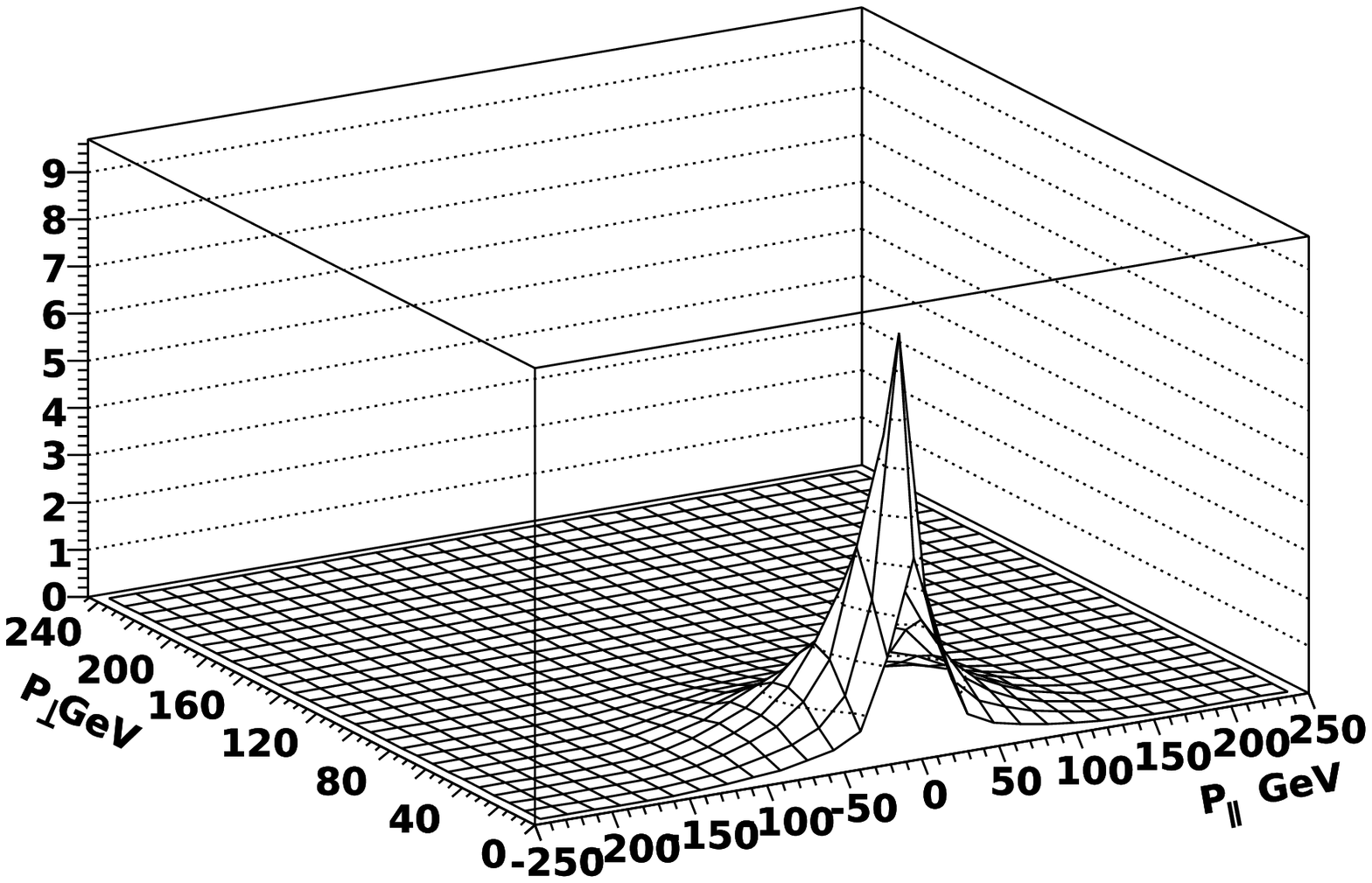}\\
 \includegraphics[scale=0.23]{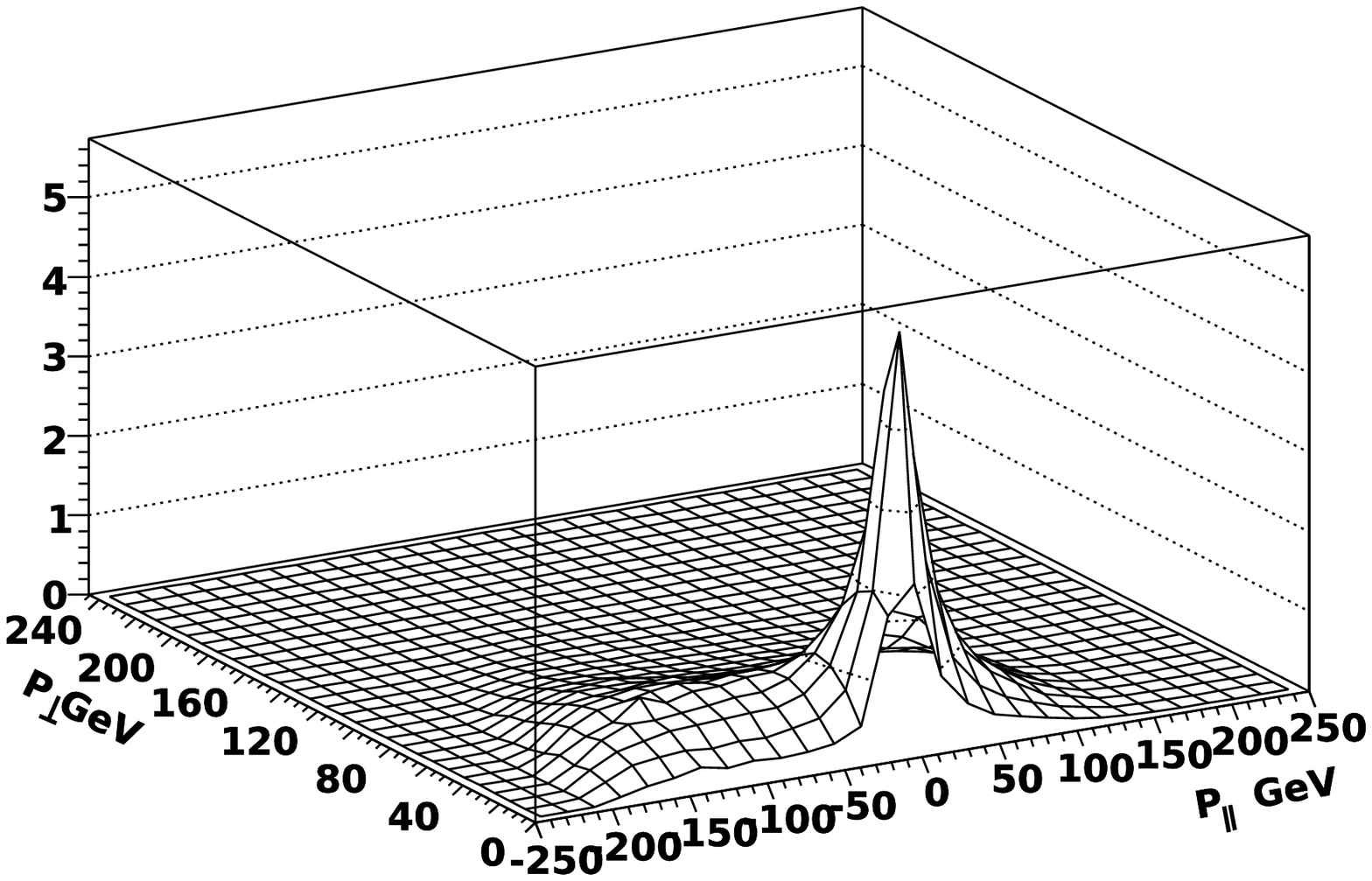}
 \includegraphics[scale=0.23]{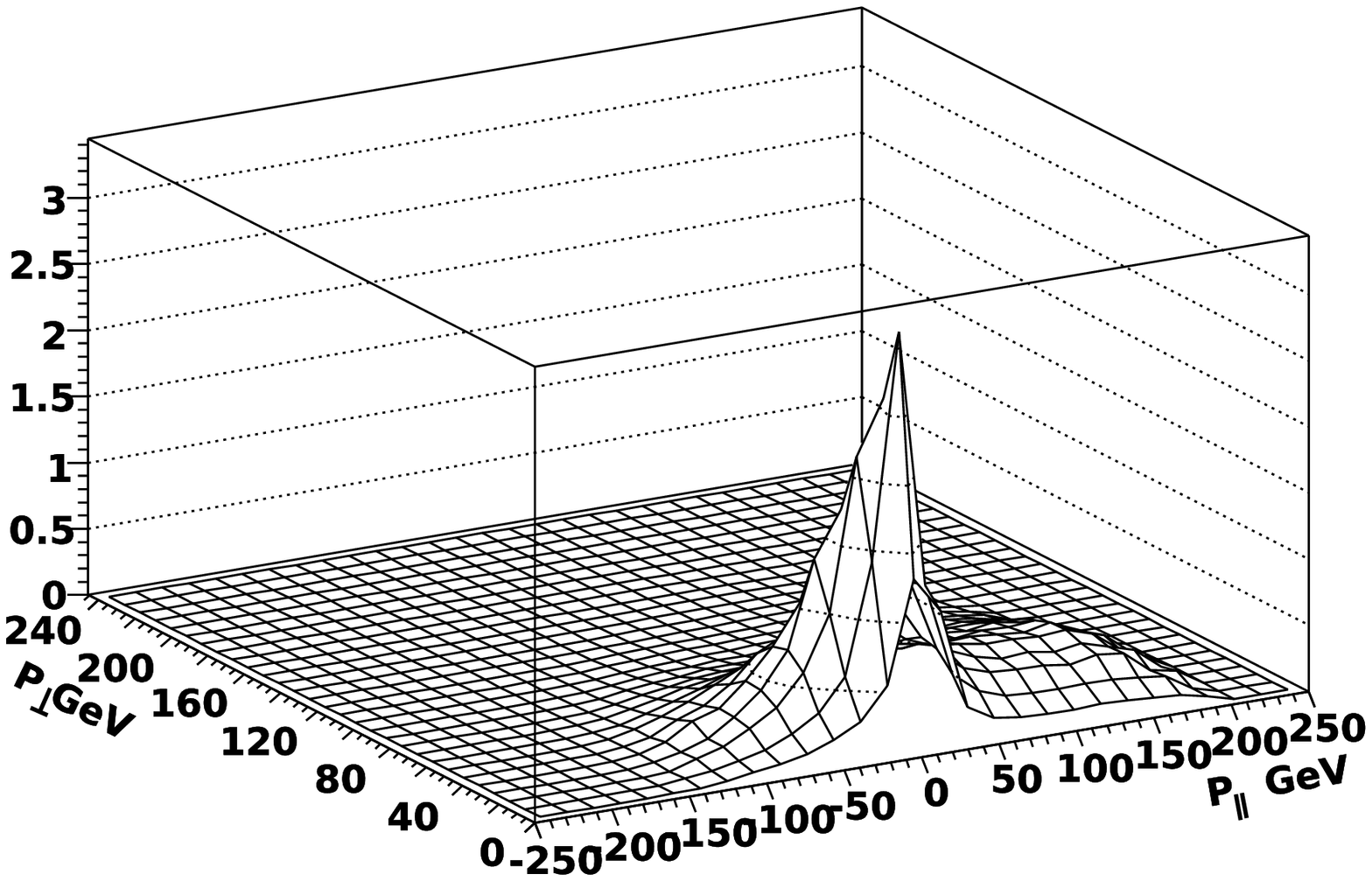}
\caption{ The distributions of muons calculated with "realistic"
spectra distribution.
        {\small Upper plots - $\gamma_-\gamma_-$. Lower plots -- $\gamma_+\gamma_-$.
        Left -- $\mu^-$,right -- $\mu^+$}} 
        \label{asymm_with_spec}
    \end{center}
\end{figure}

The resulting distributions of muons are presented on
Fig.~\ref{asymm_with_spec} for the case when incident electron
energies are 250~GeV and the laser parameter is $x=4.8$. They
resemble those presented in Fig.~\ref{asymm_no_tau+-} with
additional maximum at low energies. Table~\ref{tabspectra} shows the
corresponding asymmetry quantities. These values are slightly
smaller in comparison to the monochromatic case, but they are still
large enough and replicate in main features the values in
Table~\ref{tABI} with approximately the same statistical
uncertainties.
\begin{table*}[htb]
\begin{tabular}{||c|c||c|c|c|c|c|c|c|c|c|c|c|c||}\hline\hline
$p_{\bot\mu}^c$&$\gamma_{\lambda_1}\gamma_{\lambda_2}$ &$P_L^-$ &
$\delta P_L^-$& $P_L^+$ &$\delta P_L^+$& $\Delta_L$& $\delta
\Delta_L$&$P_T^-$ & $\delta P_T^-$& $P_T^+$ &$\delta P_T^+$&
$\Delta_T$& $\delta \Delta_T$\cr \hline
  10& $\,
\gamma_{-}\gamma_{-}\,$& $\, 0.365\,$ &$\, 0.31\% \,$ & $\, 0.157\,
$ & $\,0.22\,$\% &$\, +0.398\, $ & $\, 0.18\% \,$ &$\, 0.284\, $&
$\, 0.38\%\, $ &$\, 0.179\,$ &$\, 0.10\%\,$ &$\, +0.228\, $ &
$\,0.81\%\, $\cr \cline{2-14}
 &$\, \gamma_{+}\gamma_{-}\,$& $\, 0.174\,$ &$\, 0.24\% \,$ & $\, 0.338\,
$ & $\,0.08\,$\% &$\, -0.321\, $ & $\, 0.43\% \,$ &$\, 0.200\, $
&$\, 0.09\%\, $ &$\, 0.236\,$ &$\, 0.16\%\,$ &$\, -0.082\, $ &
$\, 0.42\%\, $\cr\hline\hline
 40& $\,
\gamma_{-}\gamma_{-}\,$& $\, 0.375\,$ &$\, 0.52\% \,$ & $\, 0.199\,
$ & $\,0.16\,$\% &$\,+0.308\, $ & $\, 0.65\% \,$ &$\, 0.352\, $&
$\, 0.15\%\, $ &$\, 0.268\,$ &$\, 0.14\%\,$ &$\, +0.136\, $ &
$\,0.51\%\, $\cr \cline{2-14}
 &$\, \gamma_{+}\gamma_{-}\,$& $\, 0.204\,$ &$\, 0.51\% \,$ & $\, 0.386\,
$ & $\,0.13\,$\% &$\, -0.308\, $ & $\, 0.56\% \,$ &$\, 0.278\, $
&$\, 0.14\%\, $ &$\, 0.319\,$ &$\, 0.13\%\,$ &$\, -0.067\, $ &
$\,0.82\%\, $\cr\hline\hline
 80& $\,
\gamma_{-}\gamma_{-}\,$& $\, 0.355\,$ &$\, 0.47\% \,$ & $\, 0.208\,
$ & $\,0.23\,$\% &$\, +0.263\, $ & $\, 0.88\% \,$ &$\, 0.515\, $&
$\, 0.08\%\, $ &$\, 0.449\,$ &$\, 0.06\%\,$ &$\, +0.069\, $ &
$\,0.53\%\, $\cr \cline{2-14}
 &$\, \gamma_{+}\gamma_{-}\,$& $\, 0.207\,$ &$\, 0.38\% \,$ & $\, 0.338\,
$ & $\,0.17\,$\% &$\, -0.305\, $ & $\, 0.75\% \,$ &$\, 0.467\, $
&$\, 0.06\%\, $ &$\, 0.483\,$ &$\, 0.03\%\,$ &$\, -0.017\, $ &
$\, 2.53\%\, $\cr\hline\hline
\end{tabular}
\caption{Charge asymmetry quantities for "realistic"
photon spectra, $\sqrt{s_{ee}}=500$ GeV.}
\label{tabspectra}
\end{table*}
\section{Correlative asymmetries in \boldmath{$\gamma \gamma \to \mu^+ \mu^- +\nu's$}}\label{seccorrel}

\begin{figure}[t]
\begin{center}
\includegraphics[scale=0.4]
{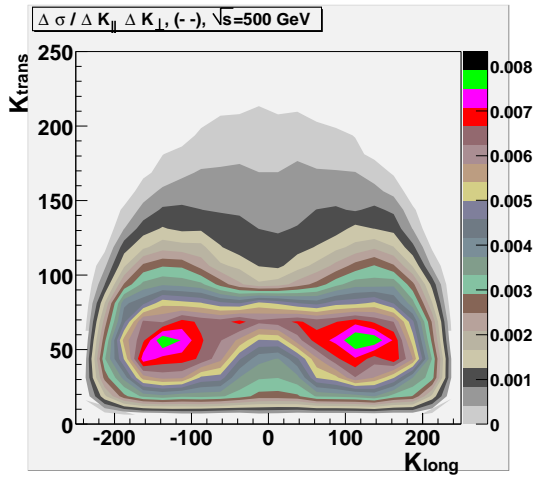}
\caption{Distribution in
$k_\|,k_\bot$, monochromatic photons, $\sqrt{s}=500$ GeV}
\label{kdistr}
\end{center}
 \end{figure}

The charge asymmetry in relative distributions of positive and
negative muons in each event can be a more useful instrument to hunt
for the New Physics (but with lower counting rates). A simple
analogy in terms00 of charge symmetric variables, is provided by
transverse momentum and invariant mass distribution.  The global
asymmetry distribution corresponds  to that in transverse momentum
while the correlative asymmetry distribution corresponds to that in
the effective $\mu^+\mu^-$ mass. The latter is sensitive to the
existence of possible resonance states, which cannot be seen in
global asymmetries. As an example we present distribution in
$\vec{k}=\vec{p}_+ + \vec{p}_-$ in its longitudinal and transverse
component, Fig.~\ref{kdistr}.

In the case of charge symmetry this distribution would be centered
around the point $(k_\|,k_\bot)=(0,0)$. This figure exhibits strong
effect of charge asymmetry.

The first problem for numerical analysis here is to find some
representative variables in 5--dimensional space of observables
$\vec{p}_+\,\vec{p}_-$. We consider three representative "natural"
dimensionless variables for $\gamma\gamma \to\mu^+\mu^-\nu\bar{\nu}$
process:
\begin{eqnarray}&& v=\dfrac{  4(p_{\bot +}^2-p_{\bot-}^2)}{  M_W^2},\quad
u=\dfrac{ 4(p_{\parallel+}^2-p_{\parallel-}^2)}{  M_W^2},\quad \nonumber \\
&&w=\dfrac{ 4(p_{\parallel+}\epsilon_+ - p_{\parallel-}\epsilon_-)}{
M_W^2}, \end{eqnarray}
where $\epsilon_{\pm}$ is the energy of the
$\mu^\pm$. The asymmetry quantities are mean values of this
quantities averaged over all events allowed by cuts. In the future
study of effects of New Physics some other variables can be more
useful.

\begin{figure}[b]
\begin{center}
\includegraphics[scale=0.28]{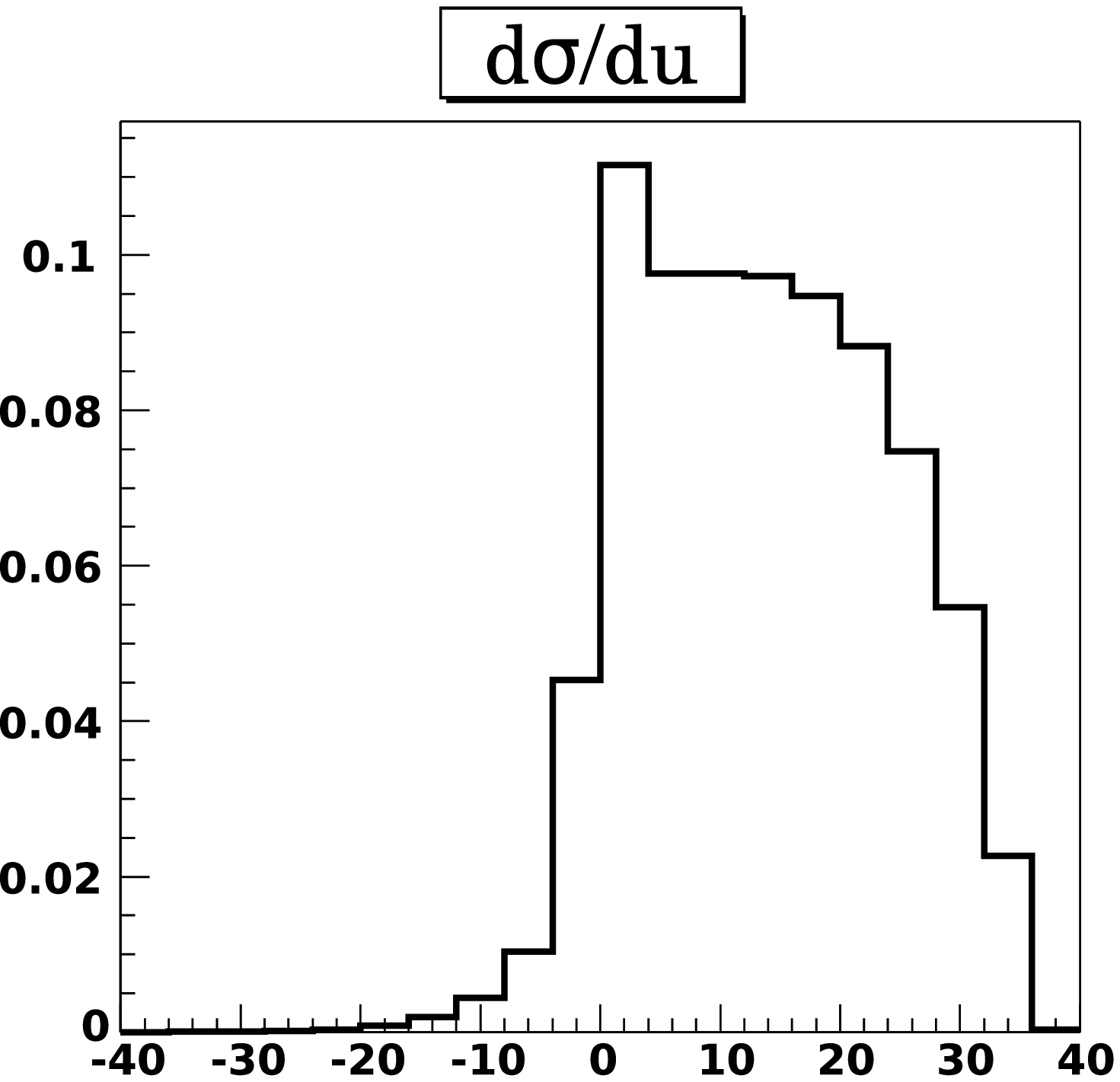}
\includegraphics[scale=0.28]{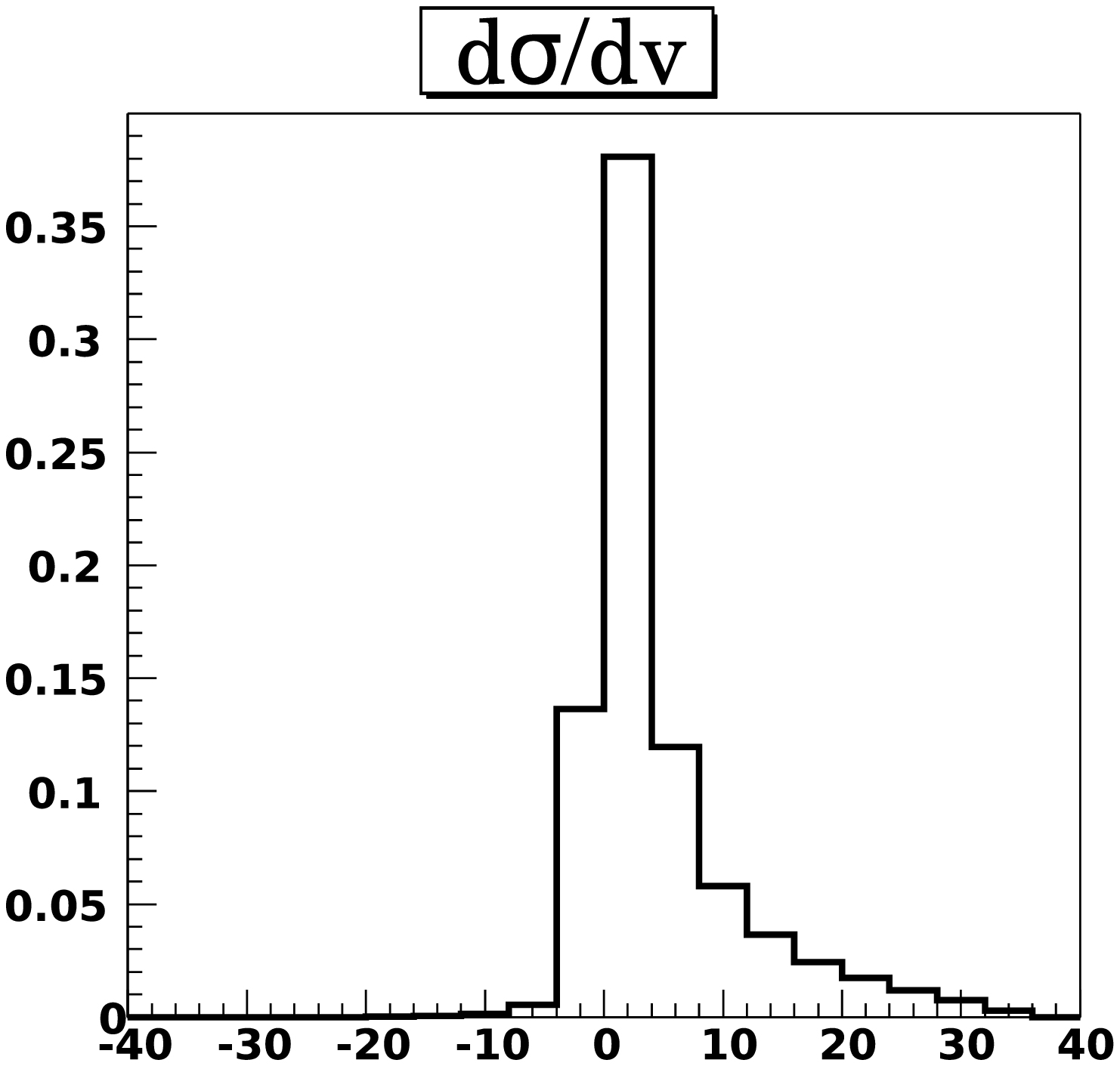}\\
\includegraphics[scale=0.3]{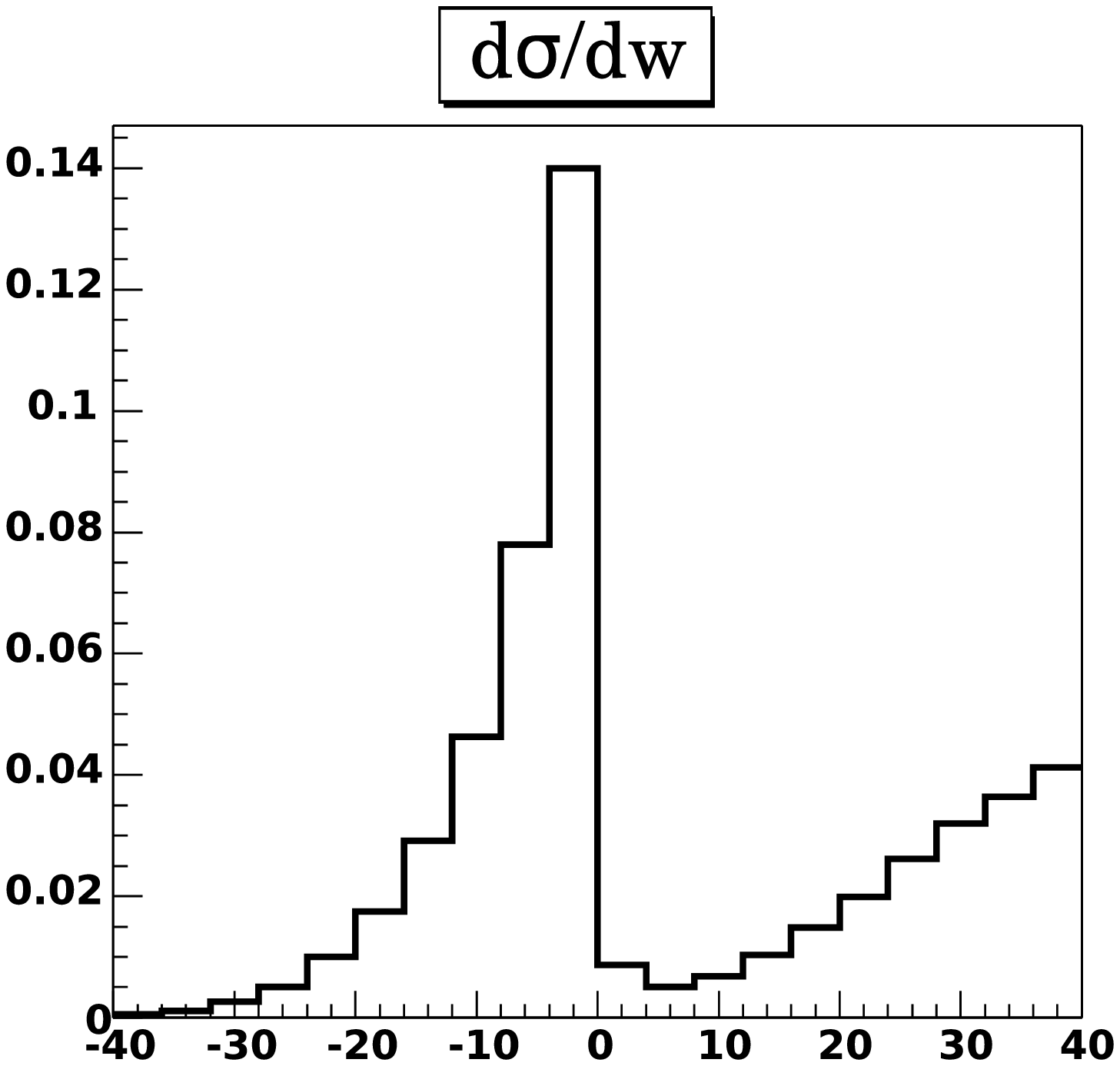}
 \caption{ Distribution in $u$ (left) and
$v$ (right) for  $\gamma_-\gamma_-$ collision. Bottom:
 Distribution in $w$ for $\gamma_-\gamma_+$ collision.
The $\gamma\gamma \to\mu^+\mu^-\nu\bar{\nu}$ process with
monochromatic photons at $\sqrt{s}=500$  GeV} \label{vvn}
\end{center}
\end{figure}

We present distributions in these variables in Fig.~\ref{vvn}.

For the $\gamma_-\gamma_-$ and $\gamma_+\gamma_+$ collisions the
distributions in the forward and backward hemispheres are identical.
For these initial photon polarizations $w=0$ while the variables $u$
and $v$ describe interesting asymmetries. Vice versa, for the
$\gamma_-\gamma_+$ collision the distributions in the forward and
backward hemispheres can be obtained from each other by the exchange
$\mu^\pm\leftrightarrow \mu^\mp$. Therefore, for these collisions
$u=v=0$ while $w$ describe the charge asymmetry, see
Figure~\ref{vvn}.

\section{Summary and outlook}\label{secdisc}

Let us enumerate main results obtained in this work.

\bu  We consider the  \emph{charge asymmetry} of leptons produced together with neutrinos in the collision of polarized photons. This charge asymmetry  is defined as the difference in the momentum distributions of the produced negatively and positively charged  leptons, and arises
because the CP conserving weak interaction vertex makes the momentum distributions strongly correlated to the initial photon polarization. This asymmetry is observable for each fixed circular polarization of at least one colliding photon.

\bu \ In particular,  we  present a detailed analysis of charge
asymmetries,  in the SM reactions $\gamma \gamma \to W^\pm \ell^\mp + \nu's$ and $\gamma\gamma\to\ell^+\ell^-+\nu's$, with polarized photons.
The method of observation of this effect,  described in detail in  the text, is based on a standard differential analysis of final state  momentum distributions of the  observed leptons with suitable applied cuts, and using well known Monte Carlo software for the generation of events.

\bu \ We suggest the method for obtaining an estimate of the lower bound for the statistical uncertainty of future experiments as given by the error  of the  Monte Carlo simulation at the anticipated number of events. We find that this uncertainty for the quantities under interest in our problem is  significantly larger than $1/\sqrt{N}$ (by a factor $3\div 5$).

\bu\ Table I shows that the statistical uncertainty in the charge asymmetry is at the level of radiative corrections. Therefore loop corrections to the differential distributions and the resulting corrections to the charge asymmetries can be safely neglected in the analysis of this type of experiments to be performed at a photon collider. One can hope to observe the effects of radiative corrections only if the luminosity would be enhanced by a factor $10\div 100$.

\bu \  Processes with intermediate  tau lepton decays  (cascade process) do also
contribute to the final state with $\ell^\pm$. We have constructed an approximation,  which describes cascade processes simply (based on the double resonant
diagrams for $W^\pm$ pair production). This approximation describes the contribution from cascade
processes to the observable charge asymmetries with high enough accuracy, within the statistical uncertainty of future experiments.

\bu \ Taking into account the cascade processes changes the charge asymmetry only weekly, the relative value of this contribution decreases at increasing values of the cut-off momentum $p_\bot^c$.

\bu \ We have further shown that the non-monochromaticity of photons at Photon Colliders diminishes the considered asymmetries, but  only weekly.

\bu \ The substantial reduction of cross sections at increasing values of the cut-off momentum $p_\bot^c$ above $M_W/2$ is to be compared with the fact that, on the contrary, the charge asymmetries are affected only slightly by $p_\bot^c$. This makes the charge asymmetry a very good candidate as optimal observable  for the discovery of New Physics effects  in the processes $\ggam\to \ell^+\ell^-+neutrals$ if, as it is expected, the scale of the New Physics is larger than $M_W$.

\acknowledgments This work is  supported by grants RFBR
08-02-00334-a and NSh-1027.2008.2. I. F. Ginzburg acknowledges
support from the \emph{Centro di Cultura Scientifica}
�\emph{Alessandro Volta}�, Landau Network office,  which allowed
a visit to INFN Sezione di Perugia, where this work was initiated.
This work was also partially supported, in the earlier stages, by
the European Contract HPMF-CT-2000-0752. K.Kanishev is supported
also by EU Marie Curie Research Training Network FLAVIAnet under
contract No. MRTN-CT-2006-035482

\end{document}